 \documentclass[10pt,preprint]{aastex}  %compact preprint style (MJA)
 \usepackage{natbib}
\def\gapprox{\lower.4ex\hbox{$\;\buildrel >\over{\scriptstyle\sim}\;$}}
\def\lapprox{\lower.4ex\hbox{$\;\buildrel <\over{\scriptstyle\sim}\;$}}

\shortauthors{ASCHWANDEN}
\shorttitle{Harmonic Resonances of Planets}

\begin{document}
%{\sl  Manuscript, accepted ... }

\title{         Self-Organizing Systems in Planetary Physics : 
		Harmonic Resonances of Planet and Moon Orbits } 

\author{        Markus J. Aschwanden$^1$}

\affil{		$^1)$ Lockheed Martin, 
		Solar and Astrophysics Laboratory, 
                Org. A021S, Bldg.~252, 3251 Hanover St.,
                Palo Alto, CA 94304, USA;
                e-mail: aschwanden@lmsal.com }

\begin{abstract}
The geometric arrangement of planet and moon orbits into a 
regularly spaced pattern of distances is the result of a 
self-organizing system. The positive feedback mechanism
that operates a self-organizing system is accomplished by 
harmonic orbit resonances, leading to long-term stable
planet and moon orbits in solar or stellar systems. 
The distance pattern of planets was originally
described by the empirical Titius-Bode law, and by a generalized 
version with a constant geometric progression factor (corresponding
to logarithmic spacing).
We find that the orbital periods $T_i$ and planet distances $R_i$ from
the Sun are not consistent with 
logarithmic spacing, but rather follow the quantized scaling 
$(R_{i+1}/R_i) = (T_{i+1}/T_i)^{2/3} = (H_{i+1}/H_i)^{2/3}$, where
the harmonic ratios are given by five dominant resonances, namely
$(H_{i+1} : H_{i}) = (3:2), (5:3), (2:1), (5:2), (3:1)$.
We find that the orbital period ratios tend to follow the quantized harmonic
ratios in increasing order. We apply this harmonic orbit resonance model
to the planets and moons in our solar system, and to the exo-planets 
of 55 Cnc and HD 10180 planetary systems. 
The model allows us a prediction of missing planets 
in each planetary system, based on the quasi-regular self-organizing 
pattern of harmonic orbit resonance zones. We predict 7 (and 4) missing 
exo-planets around the star 55 Cnc (and HD 10180).
The accuracy of the predicted planet and moon distances amounts 
to a few percents.  All analyzed systems are found to have 
$\approx 10$ resonant zones that can be occupied with planets 
(or moons) in long-term stable orbits. 
\end{abstract}

\section{	INTRODUCTION			}

Johannes Kepler was the first to study the distances of
the planets to the Sun and found that the inner radii of
regular geometric bodies (Platon's polyhedra solids) approximately
match the observations, which he published in his famous
{\sl Mysterium Cosmographicum} in 1596. An improved
empirical law was discovered by J.B. Titius in 1766, and
it was made prominent by Johann Elert Bode (published in 1772), 
known since then as the {\sl Titius-Bode law}:
\begin{equation}
	R_n = \left\{ \begin{array}{ll}
		0.4	& \mbox{for $n=1$} \\ 
		0.3 \times 2^{n-2} + 0.4 & \mbox{for $n=2,...,10$}
		\end{array}
		\right.
\end{equation}
Only the six planets from Mercury to Saturn were known at that time.
The asteroid belt, represented by the largest asteroid body 
{\sl Ceres} (discovered in 1801), part of the so-called
``missing planet'' (Jenkins 1878; Napier et al.~1973; Opik 1978),
was predicted from the Titius-Bode law, as well as the outer
planets Uranus, Neptune, and Pluto, discovered in 1781, 1846, and 1930,
respectively. Historical reviews of the Titius-Bode law can be found 
in Jaki (1972a,b), Ovenden (1972, 1975), Nieto (1972), Chapman (2001a,b), 
and McFadden et al.~(1999, 2007). 

Noting early on that the original Titius-Bode law breaks
down for the most extremal planets (Mercury at the inner side,
and Neptune and Pluto at the outer side),
numerous modifications were proposed: such as a 4-parameter
polynomial (Blagg 1913; Brodetsky 1914); the Schr{\"o}dinger-Bohr
atomic model with a scaling of $R_n \propto n(n+1)$, where
the quantum-mechanical number $n$ is substituted by the planet 
number (Wylie 1931; Louise 1982; Scardigli 2007a,b); 
a geometric progression by a constant factor 
(Blagg 1913; Nieto 1970; Dermott 1968, 1973; Armellini 1921; 
Munini and Armellini 1978; Badolati 1982; Rawal 1986, 1989;
see compilation in Table 1); fitting an exponential distance law
(Pletser 1986, 1988);
the introduction of additional planets (Basano and Hughes 1979),
applying a symmetry correction to the Jupiter-Sun system
(Ragnarsson 1995); tests of random statistics 
(Dole 1970; Lecar 1973; Dworak and Kopacz 1997; 
Hayes and Tremaine 1998; Lynch 2003; Neslusan 2004; Cresson 2011; 
Pletser 2017); self-organization of atomic patterns
(Prisniakov 2001), standing waves in the solar system formation 
(Smirnov 2015), or the Four Poisson-Laplace theory of gravitation 
(Nyambuya 2015).

Also the significance of the Titius-Bode law for predicting
the orbit radii of moons around Jupiter or Saturn was recognized
early on (Blagg 1913; Brodetsky 1914; Wylie 1931; Miller 1938a,b;
Todd 1938; Cutteridge 1962; Fairall 1963; Dermott 1968; Nieto 1970;
Rawal 1978; Hu and Chen 1987), or the prediction of a
trans-Neptunian planet ``Eris'' (Ortiz et al.~2007; 
Flores-Gutierrez and Garcia-Guerra 2011; Gomes et al. 2016),
while more recent usage of the Titius-Bode law is made to 
predict the distances of exo-planets to their central star 
(Cuntz 2012; Bovaird and Lineweaver 2013, Bovaird et al.~2015; 
Poveda and Lara 2008; 
Lovis et al. 2011; Qian et al. 2011; Huang and Bakos 2014),
or a planetary system around a pulsar (Bisnovatyi-Kogan 1993).  

Physical interpretations of the Titius-Bode Law involve the
accumulation of planetesimals, rather than the creation of 
enormous proto-planets and proto-satellites (Dai 1975, 1978). 
N-body (Monte-Carlo-type) computer simulations of the 
formation of planetary stystems were performed, which could 
reproduce the regular orbital spacings of the Titius-Bode law 
to some extent (Dole 1970; Lecar 1973; Isaacman and Sagan 1977; 
Prentice 1977; Estberg and Sheehan 1994), or not (Cameron 1973). 
Some theories concerning the Titius-Bode law
involve orbital resonances in planetary system formation,
starting with an early-formed Jupiter which produces a runaway
growth of planetary embryos by a cascade of harmonic resonances
between their orbits (e.g., Goldreich 1965; Dermott 1968;
Tobett et al. 1982; Patterson 1987; Filippov 1991).
Alternative models involve the self-gravitational instability
in very thin Keplerian disks (Ruediger and Tschaepe 1988; Rica 1995),
the principle of least action interaction (Ovenden 1972; 
Patton 1988), or scale-invariance of the disk that produces
planets (Graner and Dubrulle 1994; Dubrulle and Graner 1994).
Analytical models of the Titius-Bode law
have been developed in terms of hydrodynamics in thin disks
that form rings (Nowotny 1979; Hu and Chen 1987), 
periodic functions with Tschebischeff polynomials (Dobo 1981), 
power series expansion (Bass and Popolo 2005), and 
the dependence of the regularity parameter on the central
body mass (Georgiev 2016).

The previous summary reflects the fact that we still 
do not have a physical model that explains the empirical
Titius-Bode law, nor do we have an established
quantitative physical model that predicts the exact 
geometric pattern of the planet distances to the central star,
which could be used for searches of exo-solar planets or 
for missing moons around planets. In this Paper we investigate a 
physical model that quantitatively explains the distances of the
planets from the Sun, based on the most relevant harmonic 
resonances in planet orbits, which provides a more accurate 
prediction of planet distances than the empirical Titius-Bode law, 
or its generalized version with a constant geometric progression 
factor. This model appears to be universally applicable, to the 
planets of our solar system, planetary moon systems, Saturn-like 
ring systems, and stellar exo-planetary systems. 

A novel approach of this study is the interpretation of 
harmonic orbit resonances in terms of a self-organization system
(not to be confused with self-organized criticality systems,
Bak et al.~1987; Aschwanden et al.~2016). 
The principle of self-organization is a mechanism that creates
spontaneous order out of initial chaos, in contrast to random
processes that are governed by entropy. A self-organizing
mechanism is spontaneously triggered by random fluctuations,
is then amplified by a positive feedback mechanism, and produces
an ordered structure without any need of an external control agent. 
The manifestation of a self-organizing mechanism is often a 
regular geometric pattern with a quasi-periodic structure in space,
see various examples in Fig.~1. The underlying physics can involve
non-equilibrium processes, magneto-convection, plasma turbulence,
superconductivity, phase transitions, or chemical reactions.
The concept of self-organization has been applied to
solid state physics and material science (M\"uller and Parisi 2015),
laboratory plasma physics (Yamada 2007, 2010; Zweibel and Yamada 2009); 
chemistry (Lehn 2002), sociology (Leydesdorff 1993), 
cybernetics and learning algorithms (Kohonen 1984; Geach 2012), 
or biology (Camazine et al.~2001).
In astrophysics, self-organization has been applied to 
galaxy and star formation (Bodifee 1986; Cen 2014), 
astrophysical shocks (Malkov et al.~2000), 
accretion discs (Kunz and Lesur 2013),
magnetic reconnection (Yamada 2007, 2010; Zweibel and Yamada 2009); 
turbulence (Hasegawa 1985), 
magneto-hydrodynamics (Horiuchi and Sato 1985);
planetary atmosphere physics (Marcus 1993);
magnetospheric physics (Valdivia et al.~2003; Yoshida et al.~2010),
ionospheric physics (Leyser 2001),
solar magneto-convection (Krishan 1991; Kitiashvili et al.~2010), 
and solar corona physics (Georgoulis 2005; Uzdensky 2007).
Here we apply the concept of self-organization to the solar system, 
planetary moon systems, and exo-planet systems, based on the 
physical mechanism of harmonic orbit resonances.   

The plan of the paper is an analytical derivation of the
harmonic orbit resonance model (Section 2), an application to 
observed data of our solar system planets, the moon systems of 
Jupiter, Saturn, Uranus, and Neptune, and two exo-planet systems 
(Section 3), a discussion in the context of previous work (Section 4), 
and final conclusions (Section 5).  

\section{	THEORY   			 	} 

\subsection{	The Titius-Bode Law			}

Kepler's third law can directly be derived from the equivalence
of the kinetic energy of a planet, $E_{kin}=(1/2) m_P\ v^2$, with
the gravitational potential energy, $E_{pot}=\Gamma M_{\odot} m_P / R$,
which yields the scaling between the mean planet velocity $v$ and 
the distance $R$ of the planet from the Sun,
\begin{equation}
	v \propto R^{-1/2} \ ,
\end{equation}
and by using the relationship of the mean velocity, $v=2 \pi R / T$,
yields 
\begin{equation}
	R \propto T^{2/3} \ ,
\end{equation}
which is the familiar third Kepler law, where $\Gamma$ is the 
universal gravitational constant, $m_P$ is the planet mass, 
$M_{\odot}$ the solar mass, and $T$ is the time period of 
a planet orbit. 

The empirical Titius-Bode law (Eq.~1) tells us that there is a 
regular spacing between the planet distances $R$ and the orbital 
periods $T$, which predicts a distance ratio of $\approx 2$
for subsequent planets $(n+1)$ and $(n)$, in the asymptotic limit 
of large distances, $R \gg 0.4$ AU,  
\begin{equation}
	{R_{n+1} \over R_n} =
	{0.3 \times 2^{n-1}+0.4 \over 0.3 \times 2^{n-2}+0.4}
	\approx 2 \ .
\end{equation}
According to Kepler's third law (Eq.~3), this would imply an 
orbital period ratio of
\begin{equation}
	{T_{n+1} \over T_n} = 
	\left( {R_{n+1} \over R_n} \right)^{3/2}
	\approx 2^{3/2} \approx 2.83 \ .
\end{equation}
Thus, the Titius-Bode law predicts a non-harmonic ratio for the
orbital periods, which is in contrast to celestial mechanics models 
with harmonic resonances (Peale 1976). In the following we will also 
see that the assumption of a logarithmic spacing in planet distances is
incorrect, which explains the failure of the original
Titius-Bode law for the most extremal planets in our solar system
(Mercury, Neptune, Pluto).

\subsection{	The Generalized Titius-Bode Law 	 	}

A number of authors modified the Titius-Bode law in terms of a
geometric progression with a constant factor $Q$ between subsequent
planet distances, which was called the generalized Titius-Bode law,
\begin{equation}
	{R_{n+1} \over R_n} = Q \ ,
\end{equation}
which reads as
\begin{equation}
	R_n = R_3 \times Q^{n-3}, \quad i=1,...,n \ ,
\end{equation}
if the third planet $(i=3)$, our Earth, is used as the reference 
distance $R_3=1$ astronomical unit (AU). If we apply Kepler's third 
law again (Eq.~5), we find an orbital period ratio $q$ that is
related to the distance ratio $Q$ by $q=Q^{3/2}$, 
\begin{equation}
	q = {T_{n+1} \over T_n} = 
	\left( {R_{n+1} \over R_n} \right)^{3/2}
	= Q^{3/2} \ . 
\end{equation}
Applying this scaling law to the empirical factors $Q$ found by
various authors, we find distance ratios in the range of
$Q=1.26-2.00$ (Table 1, column 1) for geometric progression factors,
and $q=1.41-2.82$ for orbital period ratios (Table 1, column 2).
None of those time period ratios matches a low
harmonic ratio $q$, such as (3:2)=1.5, (2:1)=2, or (3:1)=3.   
To our knowledge, none of the past studies found a relationship
between the empirical Titius-Bode law and harmonic periods,
as it would be expected from the viewpoint of harmonic
resonance interactions in celestial mechanics theory,
as applied to orbital resonances in the solar system 
(Peale 1976), or to the formation of the Cassini division 
in Saturn's ring system (Goldreich and Tremaine 1978; Lissauer 
and Cuzzi 1982). Moreover, the generalized Titius-Bode law
assumes logarithmically
spaced planet distances, quantified with the constant geometric
progression factor $Q$ (Eq.~6), which turns out to be incorrect
for individual planets, but can still be useful as a simple
strategy to estimate the location of missing moons and exo-planets
(Bovaird and Lineweaver 2013; Bovaird et al.~2015).
	
\subsection{	Orbital Resonances 			}

Orbital resonances tend to stabilize long-lived orbital
systems, such as in our solar system or in planetary moon systems.
Computer simulations of planetary systems have demonstrated
that injection of planets into circular orbits tend to produce
dynamically unstable orbits, unless their orbital periods settle
into harmonic ratios, also called commensurabilities (for reviews,
see, e.g., Peale 1976; McFadden 2007). For instance, the first
three Galilean satellites of Jupiter (Io, Europa, Ganymede)
exhibit a resonance (known already to Laplace 1829),
\begin{equation}
	\nu_1 - 3 \nu_2 + 2 \nu_3 \approx 0 \ ,
\end{equation}
where the frequencies $\nu_i=1/T_i$ correspond to the reciprocal
orbital periods $T_i$, with $T_1=1.769$ days for Io, $T_2=3.551$
days for Europa, and $T_3=7.155$ days for Ganymede), fullfilling
the resonance condition (Eq.~9) with an accuracy of order
$\approx 10^{-5}$, and using more accurate orbital periods 
even to an accuracy of $\approx 10^{-9}$ (Peale 1976). 

Our goal is to predict the planet distances $R$ from the Sun
based on their most likely harmonic orbital resonances.
Some two-body resonances of solar planets are mentioned in
the literature, such as the resonances (5:2) for the
Jupiter-Saturn system, (2:1) for the Uranus-Neptune system,
(3:1) for the Saturn-Uranus system, and (3:2) for the
Neptune-Pluto system (e.g., see review of Peale 1976). 
If we find the most likely resonance between two neighbored
planet orbits with periods $T_i$ and $T_{i+1}$, we can apply
Kepler's third law $R \propto T^{2/3}$ to predict the
relative distances $R_i$ and $R_{i+1}$ of the planets
from the Sun, which allows us also to test the Titius-Bode law
as well as the generalized Titius-Bode law.

Orbital resonances in the solar system are all found for small
numbers of integers, say for harmonic numbers in the range of
H=1 to H=5 (e.g., see Table 1 in Peale 1976). If we consider
all possible resonances in this number range, we have nine
different harmonic ratios, which includes $(H_{i} : H_{i+1})$ =
(5:4), (4:3), (3:2), (5:3), (2:1), (5:2), (3:1), (4:1), (5:1),
sorted by increasing ratios $q=(H_{i+1}/H_i)$, as shown in Fig.~2a.
The harmonic ratios vary in the range of $q=[1.2,...,5]$ (Fig.~2a).
The related planet distance ratios $Q$ can be obtained from
Kepler's third law, which produces ratios of $Q=q^{2/3}$
(Eq.~8), which yields a range of $Q=[1.129, ...., 2.924]$
(Fig.~2b). This defines possible distance ratios 
$Q=R_{i+1}/R_i$ between neighbored planets varying by a factor
of three, which is clearly not consistent with a single constant, 
as assumed in the generalized Titius-Bode law. 

The frequency of strongest gravitational interaction between
two neighbored planets is given by the time interval $t_{conj}$ 
between two subsequent planet conjunctions, which is defined by 
the orbital periods $T_1$ and $T_2$ as
\begin{equation}	
	{1 \over t_{conj}} = {1 \over T_1} - {1 \over T_2} \ ,
	\qquad {\rm for\ } T_2 > T_1 \ .
\end{equation}
We see that the conjunction time $t_{conj}$ approaches infinity 
in the case of two orbits in close proximity ($T_2 \gapprox T_1$), 
while it becomes largest for $T_2 \gg T_1$. The conjunction time
is plotted as a function of the harmonic ratios in Fig.~2c, which
clearly shows that the conjunction times decreases (Fig.~2c) with 
increasing harmonic ratios (Fig.~2a). Since the gravitational
force decreases with the square of the distance $\Delta R = (R_2-R_1)$, 
i.e., $F_{grav} \propto m_1 m_2/\Delta R^2$, neighbored planet pairs 
matter more for gravitational interactions, such as for stabilizing 
resonant orbits, than remote planets. On the other side,
if the planets are too closely spaced, they have similar orbital
periods and the conjunction times increases, which lowers the chance
for gravitational stabilization interactions. So, there is a trade-off
between these two competing effects that determines which relative 
distance is
at optimum for maintaining stable long-lived orbits. We will see
in the following that the ``sweet spot'' occurs for harmonic ratios
between $q=3/2=1.5$ and $q=3/1=3.0$, a range that includes only
five harmonic ratios, namely (3:2), (5:3), (2:1), (5:2), (3:1),
which are marked with hatched areas in Fig.~2.

\subsection{	The Harmonic Orbit Resonance Model  	}

We investigate now a quantitative model that can fit the
distances of the planets from the Sun, which we call the
harmonic orbit resonance model.
The basic assumption in our model is that a two-body resonance
exists between two neighbored planets in stable long-term orbits, 
which can be defined by the resonance condition (similar to Eq.~9),
\begin{equation}
	(H_i \nu_i - H_{i+1} \nu_{i+1}) \propto \omega_{i,i+1} \ ,
\end{equation}
where $H_i$ and $H_{i+1}$ are the (small) integer numbers of a 
harmonic ratio $q_{i,i+1}=(H_{i+t}/H_i)$, and $\nu_1$ and $\nu_2$
are the frequencies of the orbital periods, $\nu_i=1/T_i$,
which can be expressed as, 
\begin{equation}
	1 - {H_{i+1} \over H_i} \left({T_i \over T_{i+1}}\right) 
	= \omega_{i,i+1} \ ,
\end{equation}
where $\omega_{i,i+1}$ is the residual that remains from
unaccounted resonances from possible third or more bodies 
involved in the resonance. The normalization by the factor 
$H_i \nu_i$ in (Eq.~11) serves the purpose to make the 
residual values compatible for different planet pairs.

In order to find the harmonic ratios $H_{i+1} : H_i$ 
that fulfill the resonance condition (Eq.~12), we have
to insert the orbital time periods $T_i$, $T_{i+1}$ 
into Eq.~(12) and find the best-fitting harmonic ratios
that yield a minimum in the absolute value of the 
resulting residual $|\omega_{i,i+1}|$. 
This procedure is illustrated in Fig.~3 for the 9 neighbored
planet pair systems, where we included all 9 cases 
of different harmonic ratios $(H_i : H_{i+1})$ =
(5:4), (4:3), (3:2), (5:3), (2:1), (5:2), (3:1), (4:1), (5:1),
sorted in rank number on the x-axis, while the residual value
is plotted on the y-axis. We see (in Fig.~3) that we find
solutions with a residual value in the range of  
$\omega_{i,i+1} \approx 0.005-0.06$ in each case.
The result for the 9 planet pairs shown in Fig.~3
confirms that the best-fit resonances are confined to
the small range of 5 cases between $q=3/2=1.5$ and $q=3/1=3.0$, 
namely (3:2), (5:3), (2:1), (5:2), (3:1), as marked in Fig.~2.
Thus, the result of the harmonic orbit resonance model
for our solar system (based on the smallest two-body residuals
$\omega_{min}$, Fig.~3), are the harmonic orbit resonances of
(3:2) for Neptune-Pluto, 
(5:3) for Venus-Earth, 
(5:2) for Mercury-Venus, Mars-Ceres, Ceres-Jupiter, Jupiter-Saturn,
(2:1) for Earth-Mars, Uranus-Neptune, and
(3:1) for Saturn-Uranus (Fig.~3), which agree with all previously
cited results (e.g., see review of Peale 1976). 

We investigated also the role of Jupiter, the most massive planet,
in the three-body resonance condition (by expanding Eq.~12),
\begin{equation}
	1 - {H_{i+1} \over H_i} \left({T_i \over T_{i+1}}\right) 
	  - {H_{Jup} \over H_i} \left({T_i \over T_{Jup}}\right) 
	= \omega_{i,i+1} \ ,
\end{equation}
but found identical results (yielding $H_{Jup}=0$), except for the
three-body configuration of Ceres-Mars-Jupiter, where an optimum
harmonic ratio of (2:1) was found, instead of (5:3), for two-body
systems. From this we conclude that the two-body interactions
of neighbored planet-planet systems are more important in the 
resonant stabilization of orbits than the influence of the 
largest giant planet (Jupiter), except for planet-asteroid pairs. 

Once we know the harmonic ratio for each pair of neighbored
planets, we can apply Kepler's third law to the orbital periods
$T_i$ and predict the distances $R_i$ of the planets,
\begin{equation}
	\left( {R_{i+1} \over R_i} \right) = 
	\left( {T_{i+1} \over T_i} \right)^{2/3} = 
	\left( {H_{i+1} \over H_i} \right)^{2/3} \ . 
\end{equation}
where $H_i$ and $H_{i+1}$ are small integer numbers (from 1 to 5)
that define the optimized harmonic orbit resonance ratios.  
Specifically, the five allowed harmonic time period ratios allow 
only the values $q_i=H_{i+1}/H_i =$ 1.5, 1.667, 2.0, 2.5, and 3.0. 
Applying Kepler's law, this selection of harmonic time periods
yields the 5 discrete distance ratios $Q_i=q_i^{(2/3)} =$
[1.31, 1.40, 1.59, 1.84, 2.08]. The (arithmetic) averages
of these predicted ratios are $< q > = 2.25 \pm 0.75$,
$< Q > = 1.70 \pm 0.4$. In the following we will
also refer to the extremal values, $[q_{min}, q_{max}]=[1.5, 3.0]$ and
$[Q_{min}, Q_{max}]=[1.31, 2.08]$. 

This physical model is distintcly different from the empircal 
Titius-Bode law (Eq.~1), which assumes a constant value
$q_i=2^{3/2}=2.83$ (Eq.~4), in the limit of $R \gg 0.4$ AU, 
or from its generalized form with an unquantified constant
$Q=q^{2/3}$ (Eq.~6). What is common to all three 
models is that the planet distances can be defined 
in an iterative way, e.g., $({R_{i+1}/R_i})$.
However, both the original and the generalized Titius-Bode
law are empirical relationships, rather than based on a
physical model. Moreover, both the original and the 
generalized Titius-Bode relationships assume a
logarithmic spacing of planet distances with a constant
geometric progression factor $Q$, while the harmonic orbit
resonance model predicts 5 quantized values for the 
planet distance ratios $Q_i$. 

\subsection{	Fitting the Geometric Progression Factor    }

So far we discussed three different models to describe the
distance pattern of planets to the Sun, which we quantified 
with the geometric progression factor $Q_i$, or time period 
progression factor $q_i=Q_i^{3/2}$, namely (1) 
$q_i \approx 2^{3/2}$ for the Titius-Bode law, (2) $q_i=const$ for 
the generalized Titius-Bode law, and (3) the five quantized
values $q_i=(H_{i+1}/H_i)$ of the five most dominant harmonic
ratios $(3:2), (5:3), (2:1), (5:2), (3:1)$ for the harmonic orbit resonance
model. Although we narrowed down the possible harmonic ratios
to five values, there is no theory that predicts in what order
these five values are distributed for a given number of planets 
or moons. Among the many possibile permutations (e.g.,
$5^{10} \approx 10^7$ for 10 planets), we make a model with
the simplest choice of including a first-order term $\Delta q$, 
besides the zero-order constant $q_i$,
\begin{equation}
	\left({T_{i+1} \over T_i}\right) = q_i = q_1 + (i-1) \ \Delta q \ , 
	\qquad i=1,...,n_p \ ,
\end{equation}
which simply represents a linear gradient of the time period
progression factor $q_i$ for each planet. The corresponding geometric
progression factor is according to Kepler's law (Eq.~3),
\begin{equation}
	\left( {R_{i+1} \over R_i}\right) = 
	Q_i = \left[ q_1 + (i-1) \ \Delta q \right]^{2/3} \ ,
	\qquad i=1,...,n_p \ .
\end{equation}
For a given set of observations with time periods 
$T_i, i=1,...,n_p$, we can then fit the model of Eq.~(15) by 
minimizing the residuals $|T_i^{model}/T_i^{obs}-1|$, 
in order to determine the gradient $\Delta q$.
If the geometric progression factor is a constant, as assumed in the 
generalized Titius-Bode law, the term will vahish, i.e., $\Delta q=0$,
while it will be non-zero for any other model.
We anticipate that the term $\Delta q$ will be positive,
because a negative value would reverse the planet distances
for high planet numbers. If the geometric progression factor
is monotonically increasing with the planet number, we expect
the lowest admitted harmonic value of $q_1=(3/2)=1.5$ for the
first planet, and the highest admitted harmonic value of
$q_n=(3/1)=3.0$ for the outermost planet $n$.
We will describe the results
of the data fitting in Section 3. Once we determined the
functional form of the progression factor $q_i$, we can then
easily find missing planets or moons based on the theoretical 
progression factors predicted by the model of Eq.~(16).

\subsection{	Self-Organization of Planet Distances	     }

We interpret now the evolution of the most stable planet
orbits as a self-organizing process, which produces a regular
geometric pattern that we characterize with the geometric
($Q_i$) or temporal progression factor $q_i$. A constant
factor $Q_i$ corresponds to a strictly logarithmic spacing,
because the planet distance increases by a constant factor
for each iterative planet number. The previously described
steps of the theory concern mostly the calculation of the
specific geometric pattern of the planet distances. Let us 
justify now the interpretation in terms of a self-organizing 
system.

Self-organizing systems create a spontaneous order, where
the overall order arises from local interactions between the
parts of the initially disordered system. In the case of our
solar system, the initially disordered state corresponds to
the state of the solar system formation by self-gravity,
where the local molecular cloud
condenses into individual planets that form our solar system.
The self-organization process is spontaneous and does not need
control by any external agent. In our solar system, it is the
many gravitational disturbances that interact beween all possible
orbits, and finally settle (over billions of years) into the 
most stable orbits that result from harmonic orbit resonances,
as observed in our solar system (Peale 1976). A self-organizing
system is triggered by random fluctuations, and then amplified
by a positive feedback. The positive feedback in the evolution
of planet orbits is given by the stabilizing gravitational
interactions in resonant orbits, while a negative feedback 
would occur when a gravitational disturbance pulls a planet 
out of its orbit, or during a migration phase of planets,
where marginally stable orbits can be disrupted. 
A self-organizing system is not controlled from outside, but
rather from all interacting interior parts. Thus, a 10-planet
system results from the evolution of $10 \times 10 =100$ two-body 
interactions, which obviously lead to a self-organized
equilibrium, unless large exterior disturbances occur (e.g., a
passing star or a migrating Jupiter), or if there is not 
sufficient critical mass to condense to a full planet in the 
initial phase. The asteroids represent such
an example of incomplete condensation, prevented by the gravitational
tidal forces of the nearby Jupiter. Finally, a self-organizing
system is robust and survives many small disturbances and can
self-repair substantial perturbations. Thus, a solar system,
or a moon system of a planet, appears to fulfill all of these 
general properties of a self-organizing system. 

\section{	OBSERVATIONS AND DATA ANALYSIS 		} 

\subsection{	The Planets in our Solar System		}

In Fig.~4 we show the distances $R$ of the planets to the Sun
in our solar system, as predicted with the Titius-Bode law
(Fig.~4a), juxtaposed to the so-called generalized Titius-Bode
law (Fig.~4b), and the harmonic orbit resonance model 
(Fig.~4c). Although the Titius-Bode law fits the observed
planet distances very well from Venus $(n=2)$ to Uranus $(n=8)$,
it breaks down at the most extremal ranges. For Mercury $(n=1)$, 
it would predict a distance of $R_1=0.55$ rather than the 
observed value of $R_1=0.387$, and thus it had been set ad hoc 
to a value of $R_1=0.4$ in the Titius-Bode law (Eq.~1). For Neptune
$(n=9)$ and Pluto $(n=10)$ the largest deviations occur.
For Pluto, a value of $R_{10}=77.2$ AU is predicted, while the
observed value is $R_{10}=39.48$ AU, which is an over-prediction
by almost a factor of two. In the overall, the Titius-Bode law
agrees with the observations within a mean and standard deviation 
of $R_{pred}/R_{obs}=1.18 \pm 0.31$ (Fig.~4a).

The generalized Titius-Bode law (Fig.~4b) shows a better overall
agreement of $R_{pred}/R=0.95\pm0.13$, which is a factor
of 2.4 smaller standard deviation than the Titius-Bode law.
We fitted the data with a constant progression factor
$Q=R_{n+1}/R_n$ and found a best-fit value of $Q=1.72$.
Some improvement over the original Titius-Bode law is that
there is no excessive mismatch for the nearest planet 
(Mercury, $n=1$) or the outermost planets (Neptune $n=9$ 
and Pluto $n=10$), although the Titius-Bode law fits somewhat 
better for the mid-range planets (from Venus to Uranus).

A significantly better agreement between the observed planet 
distances $R$ and model-predicted values $R_{pred}$ is achieved by 
using the harmonic ratios as determined from the resonance
condition for each planet (Eq.~14 and Fig.~3), yielding an 
accuracy of $R_{pred}/R=1.00\pm0.04$ (Fig.~4c),
which is a factor of 8 better than the original Titius-Bode
law, and a factor of 3 better than the generalized Titius-Bode 
law. The distance values predicted by the different discussed
models are compiled for all 10 planets in Table 2. 
Note that a better agreement between the model and the data 
is achieved with the quantized harmonic ratios $q_i$, rather
than using the logarithmic spacing assumed in the generalized
Titius-Bode law.

Is there an ordering scheme of the harmonic ratios $q_i$
in the sequence of planets $i=1,...,n$? From the harmonic ratios
displayed in Fig.~4c and Table 2 it becomes clear that there is
a tendency that the harmonic ratios $q_i$ increase with the
ordering number $i$ of a planet, with two exceptions out of the
10 planets. The first exception is the interval between Mercury
and Venus, where an additional planet can be inserted, while 
the second exception is the planet Neptune, which can be 
eliminated and then produces a pattern of monotonically increasing 
intervals. We apply this modification in Fig.~5 and fit then
our theoretical model with a linearly increasing progression
factor (Eq.~15, 16). We see that a linearly increasing
progression factor yields a best fit with an accuracy of
$R_{pred}/R_{obs}=0.96\pm0.04$, or 4\% (Fig.~5a). 
The best fit yields
a linear increment of $q=0.205$ and a harmonic range of
$q=1.40-3.24$ (Fig.~5b), close to the theoretially predicted range 
of $q=1.5-3.0$ for the allowed five dominant harmonic ratios.
Thus we can conclude that the self-organized pattern is
consistent with a linearly increasing progression factor,
at least for 8 out of the 10 planets. 

It is interesting to speculate
on the reason of the two mismatches. The observed harmonic 
ratio of Neptune-Pluto (3:2) and Uranus-Neptune (2:1) yields
a ratio of (3:2) $\times$ (2:1) = (3:1), which perfectly fits
the theoretical model and is one of the allowed harmonic ratios.
Therefore, Neptune occupies a stable orbit between Uranus
and Pluto, which does not match the primary progressive geometric 
pattern, but fits a secondary harmonic pattern.
Neptune might have joined the
solar system later, or survived on a stable, interleaved harmonic
orbit. It also might have to do with the crossing orbits of
Neptune and Pluto.  
The second exception is a missing planet between
Mercury and Venus, based on the regular pattern of a linearly
increasing progressive ratio $q$ (Eq.~15, 16). The observed
harmonic ratio is (5:2) between Mercury and Venus, which
can be subdivided into two harmonic ratios 
(5:3) $\times$ (3:2) = (5:2) to match our theoretical pattern
(Eq.~15, 16) of a monotonically increasing harmonic ratio.
Consequently, a planet orbit between Mercury and Venus is
expected in order to have a regular spacing, which could
have been occupied by an earlier existing planet that was
pulled out later, or this predicted harmonic resonance zone
was never filled with a planet. 

What are the predictions for a hypothetical planet outside
Pluto? The Titius-Bode law is unable to make a prediction,
because it over-predicts the distance of Pluto already by
a factor of 2, and thus an even larger uncertainty would be
expected for a trans-Plutonian planet. The
harmonic orbit resonance model, using a geometric 
mean extrapolation method, predicts a orbital period 
of $T_{n+1} \approx T_n^2/T_{n-1}=975$ yrs and a distance of
$R_{n+1}=R_n (T_{n+1}/T_n)^{2/3}=80$ AU for the next 
trans-Plutonian planet.
A known object in proximity is the Kuiper belt
which extends from Neptune (at 30 AU) out to approximately
50 AU from the Sun, which overlaps with Pluto, but not with
the predicted distance of a trans-Plutonian planet. 
Another nearby object is Eris, the most massive and second-largest
dwarf planet known in our solar system (Brown et al.~2005).
Eris has a highly eccentric orbit with a semi-major axis of
68 AU, so it is close to our prediction of $R_{n+1}=80$ AU. 
However, since there are many more dwarf planets known at 
trans-Neptunian distances, they could all be part of a major 
ring structure, like the asteroids. In addition, the
regular pattern predicted by our harmonic orbit resonance model
would require a harmonic ratio larger than $q=3$ for a planet
outside Pluto (Fig.~5b), in excess of the allowed five harmonic
ratios, which is an additional argument that trans-Plutonian
planets are not expected to have a stable orbit.

\subsection{	The Moons of Jupiter 			}

A recent compilation of planetary satellites lists 
67 moons for Jupiter (http://www.windows2universe.org/
our$\_$solar$\_$system.moons$\_$table.html). Only 7 out of the
67 Jupiter moons have a size of $D>100$ km, namely
Amalthea, Thebe, Io, Europa, Ganymede, Callisto,
and Himalia. The smaller objects with
$1 < D < 100$ km often occur in clusters in the
distance distribution, which may be fragments
that never condensed to a larger moon or may
be the remnants of collisional fragmentation.
The irregular spacing in the form of clusters makes
these small objects with $D < 100$ km unsuitable
to study regular distance patterns. Clusters of 
sub-planet sized bodies in the outer solar system
would require additional model components, which are 
not accomodated here at this time. 

We show a fit of the harmonic orbit
resonance model to the 7 largest Jupiter moons
in Fig.~6a, which matches our theoretical model
(Eqs.~15, 16) with an accuracy of $R_{pred}/R_{obs}
=1.01\pm0.02$. The regular spacing is shown in
Fig.~6b, where the 7 observed moon distances fit
a pattern with 10 iterative harmonic ratios.
The 7 moons include the 3 Galilean satellites
Io, Europa, and Ganymede that were known to 
exhibit highly accurate harmonic ratios already
by Laplace in 1829 (Eq.~9). 
The empty resonance zones ($n=3$, $n=8$, $n=9$) 
are found at distances of $R_3=270$ Mm,
$R_8=3370$ Mm, and $R_9=6100$ Mm, where we propose
to search for additional Jupiter moons.
The moon closest to Jupiter is Adrastea, with a distance
of 128.98 km, which is close to the Roche radius (Wylie 1931),
and thus no further moons are expected to be discovered
in nearer proximity to Jupiter. 

Since 7 moons fit a geometric pattern with 10 elements
with such a high accuracy of 2\% in the moon distance
(to the center of the hosting planet), we have a strong argument
that the underlying self-organizing mechanism based 
on stable harmonic orbit resonances predicts
the correct pattern of moon distances, but there
are holes in this geometric scheme that
are not filled by moons, which apparently do not
have a simple predictable pattern. Therefore,
our theoretical model has a predictability of 70\%
for the case of Jupiter moon distances, unless 
there exist some un-discovered moons with diameters
of $>100$ km, which we consider as unlikely.

\subsection{	The Moons of Saturn  			}

The same compilation of planetary satellites mentioned
above lists 62 moons for Saturn. The largest 6 moons
(Enceladus, Tethys, Dione, Rhea, Titan, Iapetur)
have a diameter of $D>400$ km and fit the harmonic
orbit resonance model with an accuracy of 
$R_{pred}/R_{obs}=0.95\pm0.01$. If we set the
same limit of $D > 100$ km as for Jupiter (Section 3.2),
we have 13 observed moons and find a best fit of 
$R_{pred}/R_{obs}=0.95\pm0.06$ (Fig.~7a). Inspecting
the distance spacing (Fig.~7b) we find that a
geometric pattern with 11 ratios fits the data best,
where two resonance zones are occupied by two moons
each, and 3 resonance zones are not occupied. 
Neverthelss, the spatial pattern fits the data in
the predicted range of harmonic ratios, $q=1.46-2.60$ (Fig.~7b).

\subsection{	The Moons of Uranus  			}

For Uranus, a total of 27 moons have been discovered so far,
of which 8 moons have a diameter of $D > 100$ km.
Seven of the 8 largest moons have a quasi-periodic
pattern, while Sycorax is located much further outside (Fig.~8).
The best-fit geometric pattern shows 12 resonant zones
(Fig.~8b), which fit the observed moon distances with
an accuracy of $R_{pred}/R_{obs}=0.97\pm0.07$ (Fig.~8a).
The range of best-fit harmonics $q=[1.50-2.71]$ (Fig.~8b) is 
close to the theoretical prediction $q=[1.5-3.0]$.

\subsection{	The Moons of Neptune  			}

A total of 14 moons have been reported for Neptune,
of which 6 have a diameter of $D>100$ km (Fig.~9),
namely Galatea, Despina, Larissa, Proteus, Triton,
and Nereid. We show the distances of these 6 moons 
to the center of Neptune in Fig.~9. The harmonic 
resonance model fits the 6 moons with an accuracy of 
$R_{prep}/R_{obs}=1.03\pm0.06$ (Fig.~9a). 
The range of best-fit harmonics is $q=[1.29-3.55]$ (Fig.~9b),
close to the theoretical prediction $q=[1.5-3.0]$.

\subsection{	Exo-Planets 				}

Five hypothetical planetary positions were measured in the
55 Cancri system, located at distances of 0.01583, 0.115, 
0.240, 0.781, and 5.77 AU from the center of the star (Cuntz 2012). 
Cuntz (2012) applied the Titius-Bode law and predicted
4 intermediate planet positions at 0.081, 0.41, 1.51, and 
2.95 AU, adding up to a 9-planet system. 
A similar prediction was made by Poveda and Lara (2008),
predicting two additional planets at 2.0 and 15.0 AU.
Applying the harmonic resonance model to this
data (Fig.~10), which yields a best fit with an accuracy of  
$R_{pred}/R_{obs}=1.01 \pm 0.07$ and predicts a total of 12 
planets with 7 unknown positions at
$R_2=0.022$, $R_3=0.031$, $R_4=0.048$, $R_5=0.077$, $R_8=0.40$, 
$R_{10}=1.46$, and $R_{11}=2.93$ AU. It is gratifying to see that
the later four positions predicted by our model, e.g.,
$R=0.077, 0.40, 1.46, 2.93$ AU, agree well with the predictions 
of Cuntz (2012), e.g., $R=0.081, 0.41, 1.51, 2.95$ AU.
Our harmonic orbit resonance model predicts in total a number
of 12 planets within the harmonic ratio range of 
$q=[1.61-3.11]$ (Fig.~10b), where 7 planets are un-discovered.

In the millisecond pulsar PSR 1257+12 system, two planet 
companions were discovered (Wolszczan and Frail 1992),
at distances of $R_1=0.36$ AU and $R_2=0.47$, for which
the Titius-Bode law has been applied to predict additional
planets (Bisnovatyi-Kogan 1993). The orbital distance ratio
is $R_2/R_1=1.31$ and implies (using Kepler's law) an orbital
period ratio of $T_2/T_1=(R_2/R_1)^{3/2}=1.50$ that exactly
corresponds to the harmonic ratio $(H_2:H_1)=(3/2)$, and thus
is fully consistent with the harmonic orbit resonance model.

The HRPS search for southern extra-solar planets discovered
seven periods in the star HD 10180 (Lovis et al.~2011), 
which can be translated
into distances of exo-planets from the central star by using
Kepler's law and are plotted in Fig.~11a. 
We find a total of 11 planets that 
fit the harmonic resonance model with an accuracy of
$R_{pred}/R_{obs}=0.97\pm0.10$ and we predict four 
undiscovered planets in the resonant rings $n=$2, 3, 5, 9  
with distances of $R_2=0.029$, $R_3=0.039$,
$R_8=0.089$, and $R_{10}=1.55$ AU. 

In the eclipsing polar HU Aqr, two orbiting giant planets
at distances of $R_1=3.6$ and $R_2=5.4$ AU were discovered
(Qian et al. 2011).
The ratio is $Q=5.4/3.6=1.50$ implies a period ratio of
$q=Q^{3/2}=1.84$ which is not close to a harmonic ratio.

The application of the Titius-Bode to exo-planet data
furnished 141 additional exoplanets in 68 multiple-exoplanet
systems (Bovaird and Lineweaver 2013). In a follow-up study,
Bovaird et al.~(2015) predicted the periods of 228 additional
planets in 151 multi-exoplanet systems. Huang and Bakos (2014)
searched in {\sl Kepler} Long Cadence data for the 97
predicted planets of Bovaird and Lineweaver (2013) in 56
of the multi-planet systems, but found only 5 planetary
candidates around their predicted periods and questioned
the prediction power of the Titius-Bode law. 

\subsection{	Statistics of Results			}

We summarize the statistics of results in Table 3. The
analyzed data set consists of our solar system, four
moon systems (Jupiter, Saturn, Uranus, and Neptune),
and two exo-planet systems. We found that each of these
7 systems consisted of $N_{res}=10-12$ resonant zones,
of which $N_{occ}=5-10$ were occupied with detected
satellites, while $N_{miss}=0-7$ resonant zones were 
not occupied by sizable moons (with diameters of $D>100$ km)
or detected exo-planets. For the two exo-planet systems
we predict 11 additional resonance zones that could
harbor planets (Table 4). 

The new result that each planet or moon system is found to have
about 10 resonant ring zones indicates some
unknown fundamental law for the maximum distance limit of 
planet or moon formation. The innermost distance is essentially
given by the Roche limit, while the outermost distance may 
be related to an insufficient mass density that is needed for 
gravitational condensation. Our empirical result predicts
a distance range of $R_{10}/R_1 \approx 130$ for a typical
satellite system with $N_{res} \approx 10$ satellites.
The relationship between the number of satellites and
the range of planet (or moon) distances is directly
connected to the variation of the progression factors,
which is summarized in Fig.~12 for all 7 analyzed systems.

\section{	DISCUSSION				}

\subsection{	Quantized Planet Spacing 		}

The distances of the planets from the Sun, as well as the
distances of the moons from their central body (Jupiter,
Saturn, Uranus, Neptune) have been fitted with the original Titius-Bode
law, using a constant geometric progression factor
with empirical values in the range of $Q=R_{i+1}/R_i=1.26-2.0$ (Table 1),
with the Schr{\"o}dinger-Bohr atomic model, $R_n \propto n (n+1) \approx n^2$,
Wylie 1931; Louise 1982; Scardigli 2007a,b), or with more
complicated polynomial functions (e.g., Blagg 1913). 

The assumption of a constant geometric progression factor in the
planet distances, which corresponds to a regular logarithmic spacing,
is apparently incorrect, based on the poor agreement with observations,
while the harmonic orbit resonance model has the following properties:
(1) It fits the planet distances with 5 quantized values (that relate
to the five dominant harmonic ratios) with a much higher accuracy 
than the Titius-Bode law and its generalized version;
(2) It is based on the physical model of harmonic orbit resonances,
and (3) disproves the assumption of a constant progression factor
(which corresponds to a logarithmic spacing). In constrast, it 
predicts variations between neighbored orbital periods from 
$q_{min}=1.5$ to $q_{max}=3.0$, which amounts to variations by
a factor of two. The harmonic orbit resonance model fits the
data best for a linearly increasing progression factor, starting
from $q_1=q_{min}=(3:2)=1.5$ for the first (innermost) planet pair, 
and ending with 
$q_n=q_{max}=(3:1)=3.0$ for the last (outermost) planet pair. From the
7 different data sets analyzed here we find the following mean values:
$\Delta q=0.16\pm0.05$ for the linear gradient of the time period
progression factor, $q_{min}=1.45\pm0.10$ for the minimum progression
factor, and $q_{max}=3.00\pm0.33$ for the maximum progression factor
(Table 3). If we use the mean value of $\Delta q=0.16$, the model
(Eq.~15) predicts the following ratios for a sample of 10 planets:
$q_1=1.50$, $q_2=1.66$, $q_3=1.82$, $q_4=1.98$, $q_5=2.14$,
$q_6=2.30$, $q_7=2.46$, $q_8=2.62$, $q_9=2.78$, $q_{10}=2.94$,
Rounding these values to the next allowed harmonic
number (which is a rational number), the following sequence of
harmonic ratios is predicted: 
$q_1=(3:2)$, $q_2=(5:3)$, $q_3=(5:3)$, $q_4=(2:1)$, $q_5=(2:1)$,
$q_6=(5:2)$, $q_7=(5:2)$, $q_8=(5:2)$, $q_9=(3:1)$, $q_{10}=(3:1)$,
which closely matches the sequence of observed harmonic resonances
(Table 2, column 4) in our solar system. 
These harmonic ratios match the observations closely, and thus 
provide an adequate description of the geometric pattern created
by the self-organizing system. 

\subsection{	 The Geometric Progression Factor 		}

Several attempts have been made to find a theoretical
physical model for the empirical Titius-Bode law.
There exists no physical model that can explain the 
mathematical function that was empirically found by Titius and Bode,
with a scaling relationship $2^{(n-2)}$, multiplied by 
an arbitrary factor and an additive constant (Eq.~1). 
One interpretation attempted to relate it to the
Schr{\"o}dinger-Bohr atomic model, $R_n \propto n (n+1) \approx n^2$,
Wylie 1931; Louise 1982; Scardigli 2007a,b), but the reason 
why it was found to fit the Titius-Bode law is simply because 
both series scale similarly for small integer numbers,
i.e., $2^n \propto 1, 2, 4, 8, 16, 32, ...$ versus
      $n^2 \propto 0, 1, 4, 9, 16, 25, ...$,
but this numerological coincidence does not imply that 
atomic physics and celestial mechanics can be understood by
the same physical mechanism, although both exhibit 
discrete quantization rules. 

Most recent studies assume that the physics behind the
Titius-Bode law is related to the accretion of mass
through collisions within a protoplanetary disk, clearing
out material in orbits with harmonic resonances, which
leads to a non-random distribution of planet orbits with
roughly logarithmic spacing (e.g., Peale 1976;
Hayes and Tremaine 1998; Bovaird and Linewaver 2013). 
Quantitative modeling of such a scenario is not easy,
because it implies Monte Carlo-type N-body simulations
of a vast number of planetesimals that interact with
$N^2$ mutual gravitational terms. Analytical solutions 
of N-body problems, as we know since Lagrange, are virtually
non-existent for $N \ge 3$. However, the configuration of
planets that we observe after a solar system life time 
of several billion years suggests that the observed
harmonic orbit resonances represent the most stable 
long-term solutions of a resonant system, otherwise
the solar system would have disintegrated long ago. 

In contrast to the generalized Titius-Bode law with
logarithmic spacing, we argue for a model with quantized 
geometric progression factors, based on the most relevant
harmonic ratios that stabilize resonant orbits.
From a statistical point of view we can understand that
there is a ``sweet spot'' of harmonic ratios in the
planet orbits that is not too small (because it would
lengthen the conjunction times and reduce the frequency
of gravitational interactions necessary for the
stabilization of orbits), and is not too large
(because the inter-planet distances at conjunction 
would be larger and the gravitational force weaker).
These reasons constrain the optimum range of dominant 
harmonic ratios, for which we found the 5 values
between (3/2) and (3/1). However, there are still 
open questions why there is a linear gradient $\Delta q$
in the orbit time (and geometric) progression factor,
and what determines the value of this gradient 
$\Delta q \approx 0.16$. The specific value of the 
gradient determines how many planets $n_p$ can exist 
in a resonant system within the optimum range of
harmonic numbers (from $q_{min}=1.5$ to $q_{max}=3$).
(Fig.~12).
Therefore, since we have no theory to predict the gradient
$\Delta q$, we have to resort to fitting of existing data 
and treat the gradient $\Delta q$ as an empirical variable.

\subsection{	Self-Organizing Systems 		}

One necessary property of self-organizing systems is the 
positive feedback mechanism. Solar granulation (Fig.~1a), 
for instance, is driven by subphotospheric convection,
a mechanism that has a vertical temperature gradient in
a gravitationally stratified layer. It is subject to the 
Rayleight-B\'enard instability, which can be described 
with three coupled differential equations, the so-called
Lorenz model (e.g., Schuster 1988). The positive feedback
mechanism results from the upward motion of a fluid along
a negative vertical temperature gradient, which cools off
the fluid and makes it sink again, leading to chaotic motion. 
In the limit
cycle, which is a strange attractor of this chaotic system,
the system dynamics develops a characteristic size of the
convection cells (approximately 1000 km for solar graulation),
which is maintained over the entire solar surface by this
self-organizing mechanism of the Lorenz model. 

For planet orbits, gravitational disturbances act most
strongly between planet pairs that have a harmonic ratio
of their orbit times, because the gravitational pull 
occurs every time at the same location for harmonic
orbits. Such repetitive disturbances that occur at
the same location into the same direction will pull the
planet with the lower mass away from its original orbit 
and make its orbit unstable. However, if there is a third 
body with another harmonic ratio at an opposite conjuction 
location, it can pull the unstable planet back into a more
stable orbit. This is a positive feedback mechanism that
self-organizes the orbits of the planets into stable
long-term configurations. A more detailed physical
description of the resonance phenomenon can be found
in the review of Peale (1976).

\subsection{	Random Pattern Test		}

A prediction of the harmonic orbit resonance model is that
the the spacing of stable planet orbits is not random, but
rather follows some quasi-regular pattern, which we 
quantified with the quantized spacing given with 
Eqs.~(15, 16). However, the matching of the observed
planet distances with the resonance-predicted pattern
is not perfect, but agrees within an accuracy of
$R_{pred}/R_{obs}=1.00\pm0.06$ in the statistical
average only (Table 3), the question may be asked whether
a random process could explain the observed spacing.
In order to test this hypothesis we performed a
Monte-Carlo simulation with 1000 random sets of planet
distances and analyzed it with the same numerical code
as we analyzed the observations shown in Figs.~(4-11). 
In Fig.~13 we show a 2D distribution of two
values obtained for each of the 1000 simulations:
The y-axis shows the standard deviation 
$|R_{model}/R_{sim} - 1|$ of the ratio of
modeled and simulated values of planet distances,
which is a measure of how well the model fits the
data; The x-axis shows the deviations
$[(q_{min}-1.5)^2 + (q_{max}-3.0)^2]^{1/2}$ 
of the best-fit progression factors $q_{min}$
and $q_{max}$ (added in quadrature), which is
a measure how close the simulated and theoretically predicted 
progression factors agree. From the 2D distribution
shown in Fig.~13 we see that the data set of Jupiter 
matches the model best and 
exhibits the largest deviation from the random 
values, while the other six 6 data sets are
all distributed at the periphery of the random values.
Nevertheless, all analyzed data sets are found to be
significantly different from random spacing of 
planet or moon distances. 

\subsection{	Exo-Planet Searches		}

The Titius-Bode law was also applied to exo-planets of
stellar systems, such as the solar-like G8 V star 55 Cnc
(Cuntz 2012; Poveda and Lara 1980), the millisecond pulsar 
PSR 1257+12 system (Bisnovatyi-Kogan 1993), the star 
HD 10180 (Lovis et al.~2011), the eclipsing polar HU Aqr
(Qian et al. 2011), and to over 150 multi-planet systems
observed with Kepler (Bovaird and Lineweaver 2013;
Bovaird et al.~2015; Huang and Bakos 2014). The search
in Kepler data, however, did not reveal much new detections
based on predictions with the generalized Titius-Bode law,
i.e., $R_n = R_1 Q^n$ (Huang and Bakos 2014). 
Although a similar iterative formulation is used in both
the generalized Titius-Bode law and the harmonic resonance 
model (Eq.~6, 7),
the geometric progression factor $Q$ is used as a free
variable individually fitted to each system in other studies
(e.g., Bovaird and Lineweaver 2013).

The question arises how the harmonic orbit resonance model
can improve the prediction of exo-planet candidates. 
The two examples shown in Figs.~(10-11) suggest that the
predicted spatio-temporal pattern can be fitted to incomplete
sets of exo-planets with almost equal accuracy as the data sets
from the (supposedly complete) data sets in our solar system.
This may be true if there are $\gapprox 5$ planets detected
per star, but it may get considerably ambiguous for smaller
sets of $\approx 2-4$ exo-planets per star. However, since
the harmonic orbit resonance model predicts a variable ratio
of the time period progression factor that fits existing data,
it should do better than the generalized Titius-Bode law with
a constant progression factor, as it was used in recent work
(Bovaird and Lineweaver 2013; Bovaird et al.~2015).

\section{		CONCLUSIONS 				}

The physical understanding of the Titius-Bode law is a long-standing
problem in planetary physics since 250 years. In this study we
interpret the quasi-regular geometric pattern of planet distances 
from the Sun as a result of a self-organizing process that acts 
throughout the life time of our solar system. The underlying 
physical mechanism is linked to the celestial mechanics of 
harmonic orbit resonances.  The results may be useful for searches 
of exo-planets orbiting around stars. Our conclusions are:

\begin{enumerate}

\item{The original form of the Titius-Bode law on distances of the
planets to the Sun is a purely empirical law and cannot be derived 
from any existing physical model, although it fits the observations
to some extent, but fails for the extremal planets Mercury, Neptune
and Pluto. The ``generalized form of the Titius-Bode law'' assumes 
a constant geometric progression factor $Q = R_{i+1}/R_i$, which
we find also to be inconsistent with the data, since the observed 
distance ratios vary in the range of $Q \approx 1.3-2.1$, corresponding
to a variation of $q = T_{i+1}/T_i \approx 1.5-3.0$ of the orbital
periods, according to Kepler's third law, $q \propto Q^{3/2}$.}

\item{The observed orbital period ratios $q_i$ of the planets 
in our solar system correspond to five harmonic ratios, 
$(H_{i+1}/H_i)=$ (3:2), (5:3), (2:1), (5:4), (3:1), which
represent the dominant harmonic orbit resonances that
self-organize the orbits in the solar system. We find that the
progression factor $q_i$ for time periods follows approximately
a linear function $q_i = q_1 + (i-1) \Delta q$, $i=1,...,n$,
which varies in the range from the smallest harmonic ratio
$q_1=(3:2)=1.5$ of the innermost planet to $q_n=(3:1)=3.0$ for
the outermost planet, with a gradient of $\Delta q = 0.16\pm0.05$.
The progression of orbital periods $T_i$ is quantized by the 
nearest dominant five harmonics, $q_i = [1.5, 1.667, 2.0, 2.5, 3.0]$.
Based on these harmonic ratios of the orbital periods we
predict the variation of the geometric progression factors as 
$Q_i = [q_1 + (i-1) \Delta q]^{2/3}$, which is also quantized,
$Q_i = q_i^{2/3}= [1.31, 1.40, 1.59, 1.84, 2.08]$.}

\item{Fitting the geometric progression factors predicted by
our harmonic orbit resonance model to observed data from our
solar system, moon systems, and exo-planet systems, we find
best agreement for Jupiter moons (with an accuracy of 2\%),
followed by the solar system (with an accuracy of 4\% for 8 out of 
the 10 planets). The other moon systems (of Saturn, Uranus, 
Neptune) and exo-planet systems (of 55 Cnc and HD 10180)
agree with a typical accuracy of 6\%-7\%. We demonstrated 
that these accuracies of predicted planet (or moon) distances are 
significantly different from randomly distributed (logarithmic)
distances (Fig.~13). The number of resonant zones for each star or 
planet amounts to $n_{res} \approx 10-12$, which is comparable with
the number of sizable detected moons (with a diameter of 
$D \ge 100$ km) in each moon system. For the exo-planet systems
we find best fits for $n_{res} \approx 10-11$ resonance zones,
which allows us to predict 7 missing exo-planets for the
star 55 Cnc, and 4 missing exo-planets for the star HD 10180.}
 
\item{We interpret the observed quasi-regular geometric patterns
of planet or moon distances in terms of a self-organizing system.
Self-organizing systems (the primoridal molecular cloud around
our Sun) create a spontaneous order (the Titius-Bode law) from
local interactions (via harmonic orbital resonances) between the 
internal parts (the planets or moons) of the initially disordered 
(solar) system. A self-organizing process is spontaneous and
does not need an external control agent. Initial fluctuations
are amplified by a positive feedback (by the gravitational interactions
that lead to long-term stable orbits via harmonic orbit resonances).
A self-organizing system is robust 
(during several billion years in our solar
system) and capable of self-repair after large disturbances (for
instance by a passing star or a migrating giant planet, thanks
to the stabilizing gravitational interactions of harmonic
orbit resonances). We find that the ordered patterns of planet
orbits is not always complete (a planet between Mecury and Venus
seems to be missing) and can display defects (the ``superfluous''
Neptune, similarly to defects in crystal growth). The predicted 
geometric pattern thus has a high statistical probablity, but is not
always perfectly created in self-organizing systems.}

\end{enumerate}

The present study can be applied in two ways: (1) predictions of
un-discovered exo-planets in other stellar systems; and 
(2) prediction of resonant orbits and
solutions with long-term stability in numerical N-body simulations. 
The validity of the
presented model could be corroborated by new discoveries of
predicted exo-planets (e.g., Boivard and Lineweaver 2013),
and by measuring the geometric scaling of gaps and rings in
simulated N-body accretion systems. The harmonic orbit resonance
model makes a very specific prediction about five harmonic orbit 
and distance ratios (rather than the logarithmic spacing assumed
in the generalized Titius-Bode law), which can be tested with
numerical simulations. Such a quantized geometric pattern is
also distinctly different from random systems 
(Dole 1970; Lecar 1973; Dworak and Kopacz 1997; 
Dworak and Kopacz 1997; Hayes and Tremaine 1998; Lynch 2003;
Neslusan 2004; Cresson 2011).

Just at the time of submission of this paper, the discovery of
the Trappist-1 planetary system was announced (Gillon et al.~2017,
Pletser and Basano 2017; Scholkmann 2017; 
Aschwanden and Scholkmann 2017). 
This unique exoplanet system harbors 7 planets
with a regular spacing that corresponds to the harmonic ratios
of $(T_{i+1}/T_i)=[1.60, 1.67, 1.51, 1.51, 1.34, 1.62]$.
It fulfills two of our three predictions: The period ratios
are close to the predicted harmonic ratios (3:2)=1.5 and
(5:3)=1.67, and the sequence starts with the lowest of the
predicted dominant resonances, but it does not continue to
the higher harmonic ratios (2:1, 5:2, 3:1). Either we are
missing additional planets further out, or the particular
spectral type of the central star (dwarf star M8V) favors
the lowest harmonic ratios, which have the shortest conjunction
times and therefore are the most stable configurations.

\bigskip
\acknowledgements
The author acknowledges the hospitality and partial support for 
two workshops on ``Self-Organized Criticality and Turbulence'' at the
{\sl International Space Science Institute (ISSI)} at Bern, Switzerland,
during October 15-19, 2012, and September 16-20, 2013, as well as
constructive and stimulating discussions (in alphabetical order)
with Sandra Chapman, Paul Charbonneau, Henrik Jeldtoft Jensen,
Lucy McFadden, Maya Paczuski, Jens Juul Rasmussen, John Rundle, 
Loukas Vlahos, and Nick Watkins.
This work was partially supported by NASA contract NNX11A099G
``Self-organized criticality in solar physics''.

\clearpage

%%%%%%%%%%%%%%%%%%%%%%%%%%% REFERENCES %%%%%%%%%%%%%%%%%%%%%%%%%%%%% 

\clearpage

%%%%%%%%%%%%%%%%%%%%%%%%%%% TABLE 1  %%%%%%%%%%%%%%%%%%%%%%%%%%%%%%%%%
\begin{deluxetable}{ccl}
\tablecaption{The values of geometric progression factors 
($Q = R_{i+1}/R_i$) and orbital period progression factors ($q = Q^{3/2}$) 
of solar system data are compiled from previous publications.}
\tablewidth{0pt}
\tablehead{
\colhead{Geometric}&
\colhead{Orbital period}&
\colhead{Reference}\\
\colhead{progression}&
\colhead{progression}&
\colhead{}\\
\colhead{factor}&
\colhead{factor}&
\colhead{}\\
\colhead{$Q_i$}&
\colhead{$q_i$}&
\colhead{}}
\startdata
2	& 2.82	&Titius (1766), Bode (1772), Miller 1938a,b, Fairall (1963) \\
1.7275	& 2.27	&Blagg (1913), Brodetsky (1914), Nieto 1970) \\
1.89	& 2.59	&Dermott (1968, 1973) \\
1.52	& 1.87	&Armellini (1921); Badolati (1982)\\
1.52	& 1.87	&Munini and Armellini (1978); Badolati (1982)\\
1.442	& 1.73	&Rawal (1986, 1989) \\
1.26	& 1.41	&Rawal (1986, 1989) \\
(1.31, 1.40, 1.59, 1.84, 2.08) &(1.5, 1.667, 2.0, 2.5, 3.0) 
		&Harmonic orbit resonance model (this work) \\
\enddata
\end{deluxetable}

%%%%%%%%%%%%%%%%%%%%%%%%%%% TABLE 2  %%%%%%%%%%%%%%%%%%%%%%%%%%%%%%%%%
\begin{deluxetable}{rlrrrrrrr}
\tablecaption{Observed orbital periods $T$ and distances $R$ of the 
planets from the Sun, predicted harmonic orbit resonances $(H1:H2)$, 
the Titius-Bode law $R_{TB}$, the generalized Titius-Bode law
$R_{GTB}$, and predictionts of the harmonic orbit resonance
model $R_{HOR}$ and ratios $R_{HOR}/R$.}
\tablewidth{0pt}
\tablehead{
\colhead{Number}&
\colhead{Planet}&
\colhead{Orbital}&
\colhead{Harmonic}&
\colhead{Distance}&
\colhead{Distance}&
\colhead{Distance}&
\colhead{Distance}&
\colhead{Ratio}\\
\colhead{}&
\colhead{}&
\colhead{period}&
\colhead{Resonance}&
\colhead{observed}&
\colhead{TB law}&
\colhead{GTB law}&
\colhead{HOR}&
\colhead{HOR}\\
\colhead{}&
\colhead{}&
\colhead{$T$ (yrs)}&
\colhead{$(H_{i+1}:H_i)$}&
\colhead{$R$ (AU)}&
\colhead{$R_{TB}$ (AU)}&
\colhead{$R_{GTB}$ (AU)}&
\colhead{$R_{HOR}$ (AU)}&
\colhead{$R_{HOR}/R$}}
\startdata
 1 & Mercury &   0.241 & (5:2) &   0.39 &   0.55 &   0.34 &   0.38 & 0.9839 \\
 2 & Venus   &   0.615 & (5:3) &   0.72 &   0.70 &   0.58 &   0.70 & 0.9701 \\
 3 & Earth   &   1.000 & (2:1) &   1.00 &   1.00 &   1.00 &   1.00 & 1.0000 \\
 4 & Mars    &   1.881 & (5:2) &   1.52 &   1.60 &   1.72 &   1.56 & 1.0249 \\
 5 & Ceres   &   4.601 & (5:2) &   2.77 &   2.80 &   2.95 &   2.76 & 0.9982 \\
 6 & Jupiter &  11.862 & (5:2) &   5.20 &   5.20 &   5.06 &   5.01 & 0.9639 \\
 7 & Saturn  &  29.457 & (3:1) &   9.54 &  10.00 &   8.69 &   9.43 & 0.9888 \\
 8 & Uranus  &  84.018 & (2:1) &  19.19 &  19.60 &  14.93 &  19.52 & 1.0171 \\
 9 & Neptune & 164.780 & (3:2) &  30.07 &  38.80 &  25.63 &  29.97 & 0.9968 \\
10 & Pluto   & 248.400 & ...   &  39.48 &  77.20 &  44.01 &  38.77 & 0.9820 \\
\enddata
\end{deluxetable}

%%%%%%%%%%%%%%%%%%%%%%%%%%% TABLE 3  %%%%%%%%%%%%%%%%%%%%%%%%%%%%%%%%%
\begin{deluxetable}{lccccrrr}
\tablecaption{Summary of analyzed planet and moon systems using the
harmonic orbit resonance model. We include only moons with a diameter
$D > 100$ km}
\tablewidth{0pt}
\tablehead{
\colhead{Central}&
\colhead{Number of}&
\colhead{Number of}&
\colhead{Occupied}&
\colhead{Missing}&
\colhead{Model}&
\colhead{Progression}&
\colhead{Orbital period}\\
\colhead{Object}&
\colhead{satellites}&
\colhead{resonant zones}&
\colhead{zones}&
\colhead{satellites}&
\colhead{accuracy}&
\colhead{factor}&
\colhead{progression}\\
\colhead{}&
\colhead{}&
\colhead{}&
\colhead{}&
\colhead{}&
\colhead{}&
\colhead{gradient}&
\colhead{factor}\\
\colhead{}&
\colhead{$N_{sat}$}&
\colhead{$N_{res}$}&
\colhead{$N_{occ}$}&
\colhead{$N_{miss}$}&
\colhead{$R_{pred}/R_{obs}$}&
\colhead{$dq$}&
\colhead{$q$}}
\startdata
Sun	& 10 & 10 & 10 &  0 & $0.96\pm0.04$ & 0.205 & 1.40-3.24 \\ 
Jupiter &  7 & 10 &  7 &  3 & $1.01\pm0.02$ & 0.161 & 1.41-2.85 \\
Saturn  & 13 & 11 &  8 &  3 & $0.95\pm0.06$ & 0.114 & 1.46-2.60 \\
Uranus  &  8 & 12 &  8 &  4 & $0.97\pm0.07$ & 0.110 & 1.50-2.71 \\
Neptune &  6 & 10 &  6 &  4 & $1.03\pm0.06$ & 0.251 & 1.29-3.55 \\
55 Cnc	&  5 & 12 &  5 &  7 & $1.01\pm0.07$ & 0.137 & 1.61-3.11 \\
HD 10180&  7 & 11 &  7 &  4 & $0.97\pm0.10$ & 0.152 & 1.46-2.98 \\
\hline
mean	&    &    &    &    & $1.00\pm0.06$ & $0.16\pm0.05$ & $1.45\pm0.10$ \\
	&    &    &    &    &               &               & $3.00\pm0.33$ \\
\enddata
\end{deluxetable}

%%%%%%%%%%%%%%%%%%%%%%%%%%% TABLE 4  %%%%%%%%%%%%%%%%%%%%%%%%%%%%%%%%%
\begin{deluxetable}{lcrr}
\tablecaption{Predicted exo-planets of the star 55 Cnc and HD 10180.}
\tablewidth{0pt}
\tablehead{
\colhead{Central}&
\colhead{Number of}&
\colhead{Predicted}&
\colhead{Predicted}\\
\colhead{Object}&
\colhead{harmonic orbit}&
\colhead{distance}&
\colhead{by Cuntz (2002)}}
\startdata
55 Cnc	&	2 &     0.022 AU & ...	\\
55 Cnc	&	3 &     0.031 AU & ...	\\
55 Cnc	&	4 &	0.048 AU & ...	\\     
55 Cnc	&	5 &	0.077 AU & 0.081 AU \\     
55 Cnc	&	8 &	0.40  AU & 0.41 AU \\     
55 Cnc	&      10 &	1.46  AU & 1.51 AU \\     
55 Cnc	&      11 &	2.93  AU & 2.95 AU \\     
\hline
HD 10180&       2 &     0.029 AU & ...  \\
HD 10180&       3 &     0.039 AU & ...  \\
HD 10180&       8 &     0.089 AU & ...  \\
HD 10180&      10 &     1.55  AU & ...  \\
\enddata
\end{deluxetable}

%%%%%%%%%%%%%%%%%%%%%%%%%%% FIGURE %%%%%%%%%%%%%%%%%%%%%%%%%%%%%%%%% 

\begin{figure}
\plotone{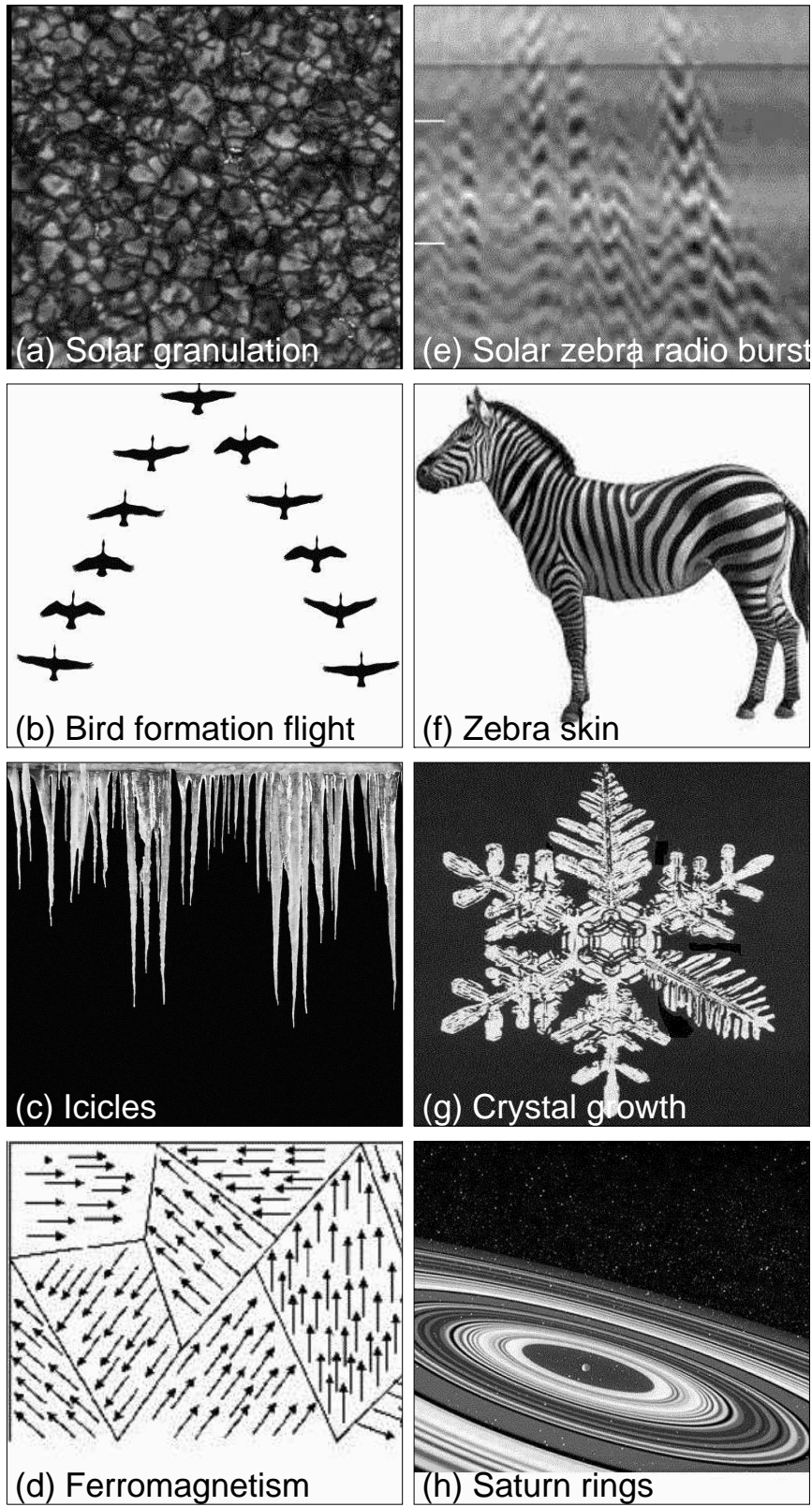}
\caption{Examples of self-organizing systems: (a) Solar granulation,
(b) Bird formation flight, (c) Icicles, (d) Ferromagnetism,
(e) Solar zebra radio bursts, (f) Zebra skin, (g) Crystal growth,
and (h) Saturn rings.}
\end{figure}

\begin{figure}
\plotone{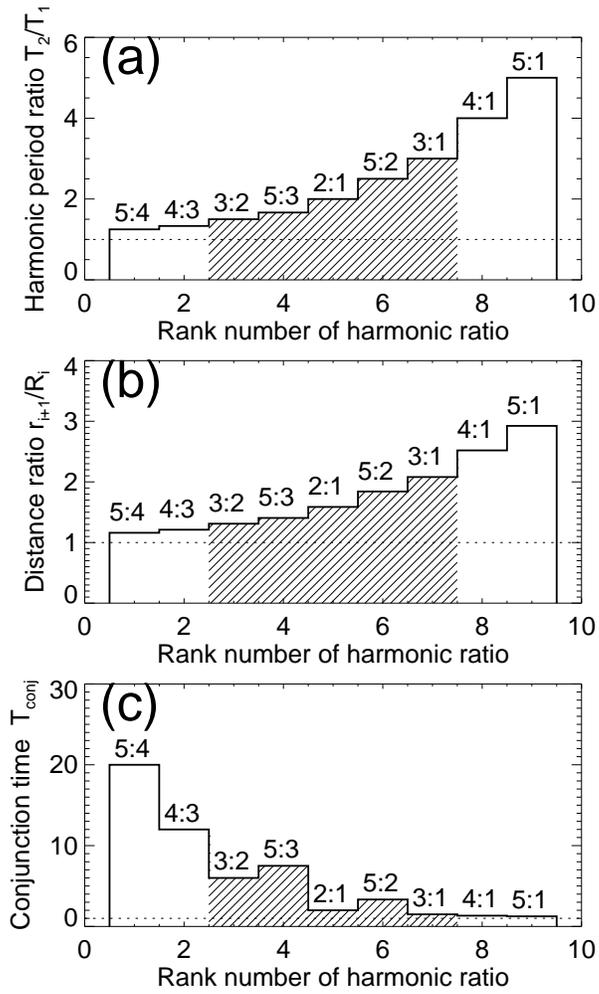}
\caption{(a) Harmonic period ratios; (b) Distance ratios of planets 
from Sun; (c) Conjunction time. The most dominant harmonic ratios
are marked with hatched line style.}
\end{figure}

\begin{figure}
\plotone{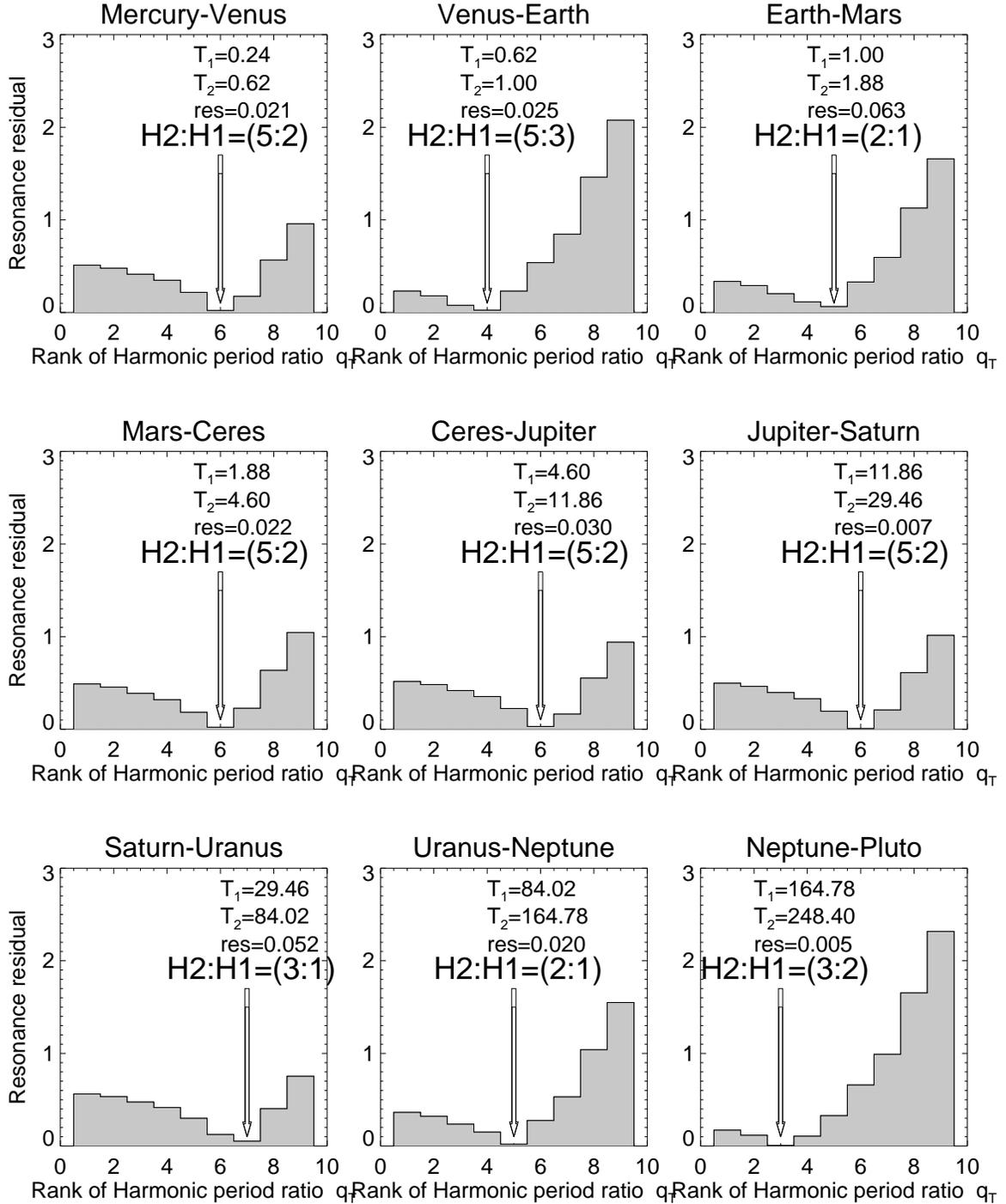}
\caption{The harmonic period ratio H2:H1 is predicted based on 
the minimum of the resonance condition residuals $\omega_{min}$ (y-axis)
as a function of the (rank number of the) harmonic period ratios 
$q_T = T_2/T_1$ (x-axis), for the 9 planet pairs (with orbital time
periods $T_1$ and $T_2$) in our solar system. The relationship
between the rank number (x-axis) and the period ratio $q_T$ is
given in Fig.~2a. Note that a minimum with a singular value of 
$\omega_{min} < 0.05$ exists in all 9 cases.}
\end{figure}

\begin{figure}
\plotone{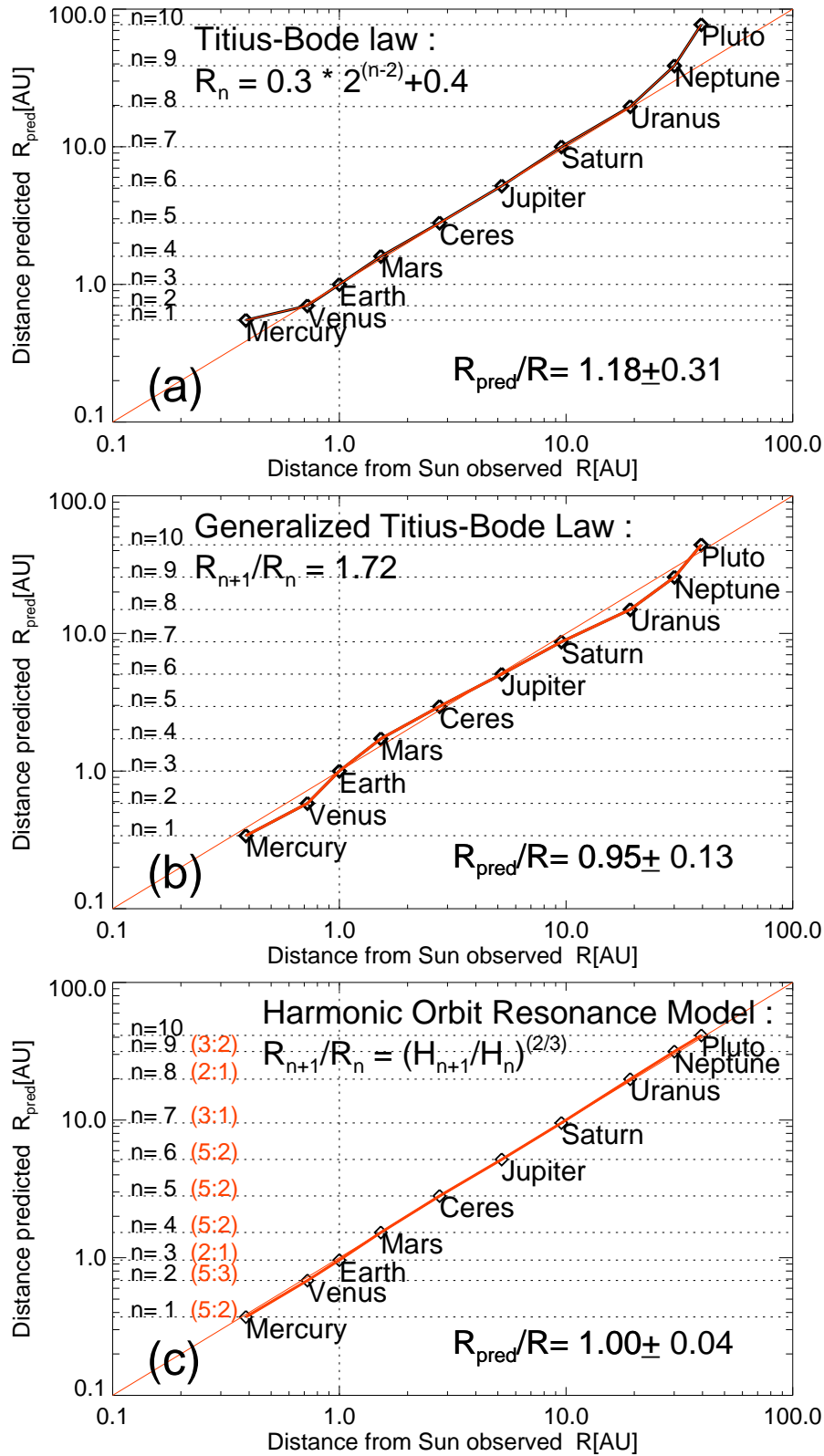}
\caption{(a) The empirical Titius-Bode law of planet 
distances from the Sun; 
(b) The generalized Titius-Bode law with a constant
progression factor $Q=1.72$;
(c) The harmonic orbit resonance model.}
\end{figure}

\begin{figure}
\plotone{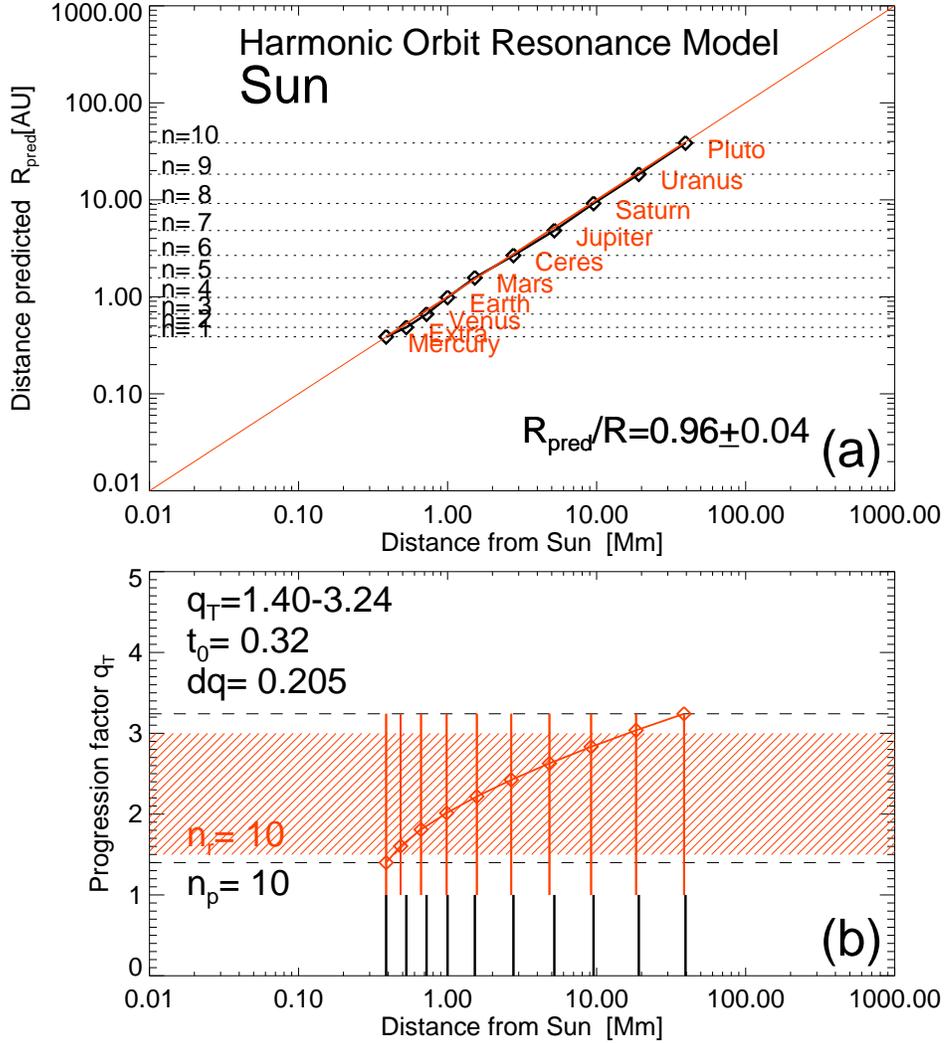}
\caption{The distances of the planets from the Sun calculated
with a harmonic orbit resonance model (a), 
for which the progression factor increases linearly 
with the orbit time, $dq=0.205$ (b). 
The observed distances are indicated with black tick marks, 
and the best-fit values with red lines and tickmarks. The
zone between the minimum and maximum progression factor is
indicated with dashed lines $q_T=1.40-3.24$, and the 
theoretically predicted range $q_T=1.5-3.0$ with a red 
hatched area. In this model two planets are modified: 
an extra planet between Mercury and Venus is inserted,
and the planet Neptune is eliminated.}
\end{figure}

\begin{figure}
\plotone{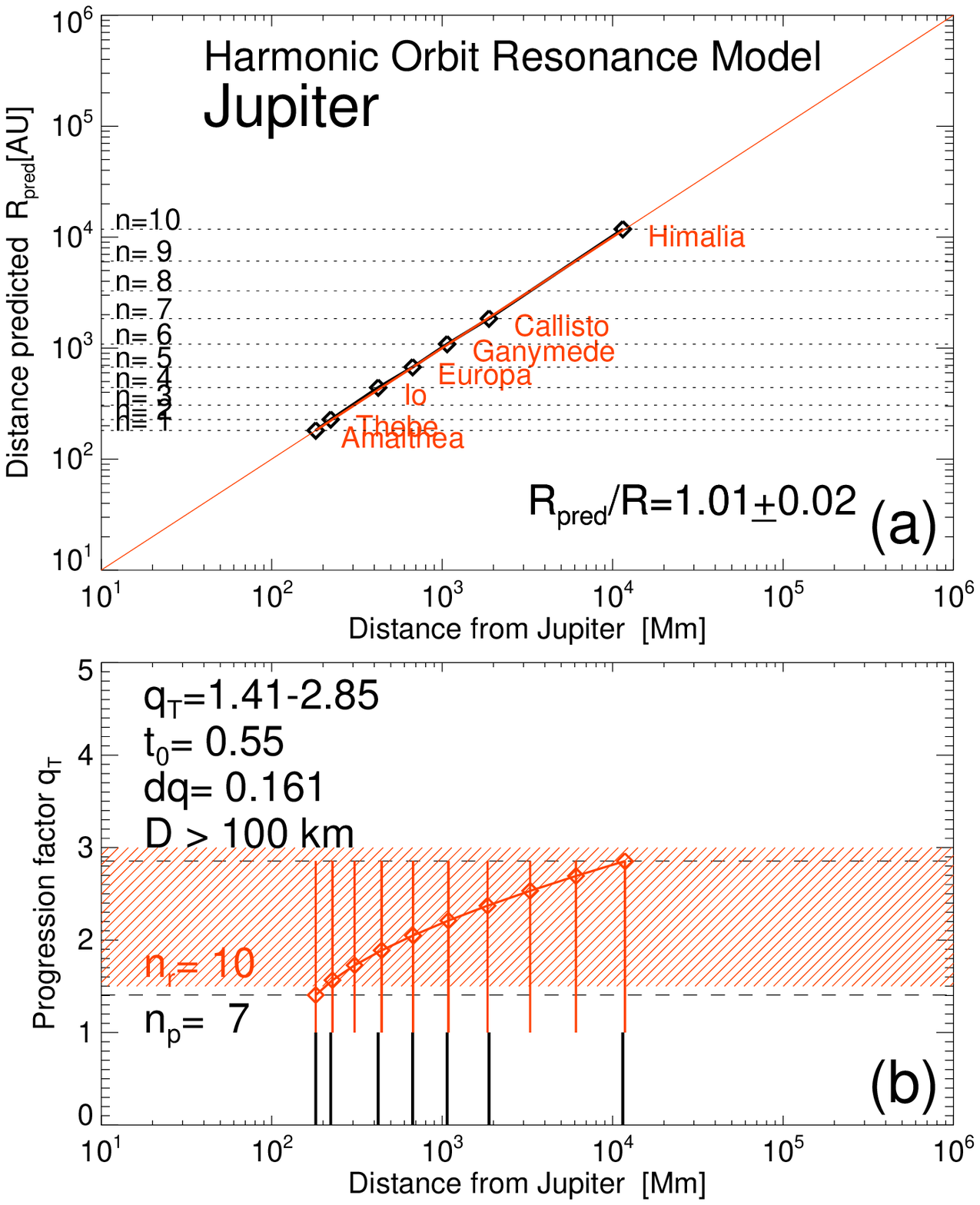}
\caption{The distances of the seven largest ($D>100$ km) 
moons from Jupiter are calculated with a harmonic
orbit resonance model (a), for which the progression factor
increases linearly with the orbit time (b). Representation
otherwise similar to Fig.~5. Note that $n=10$ resonant
zones fit $n_p=7$ observed moon distances.}
\end{figure}

\begin{figure}
\plotone{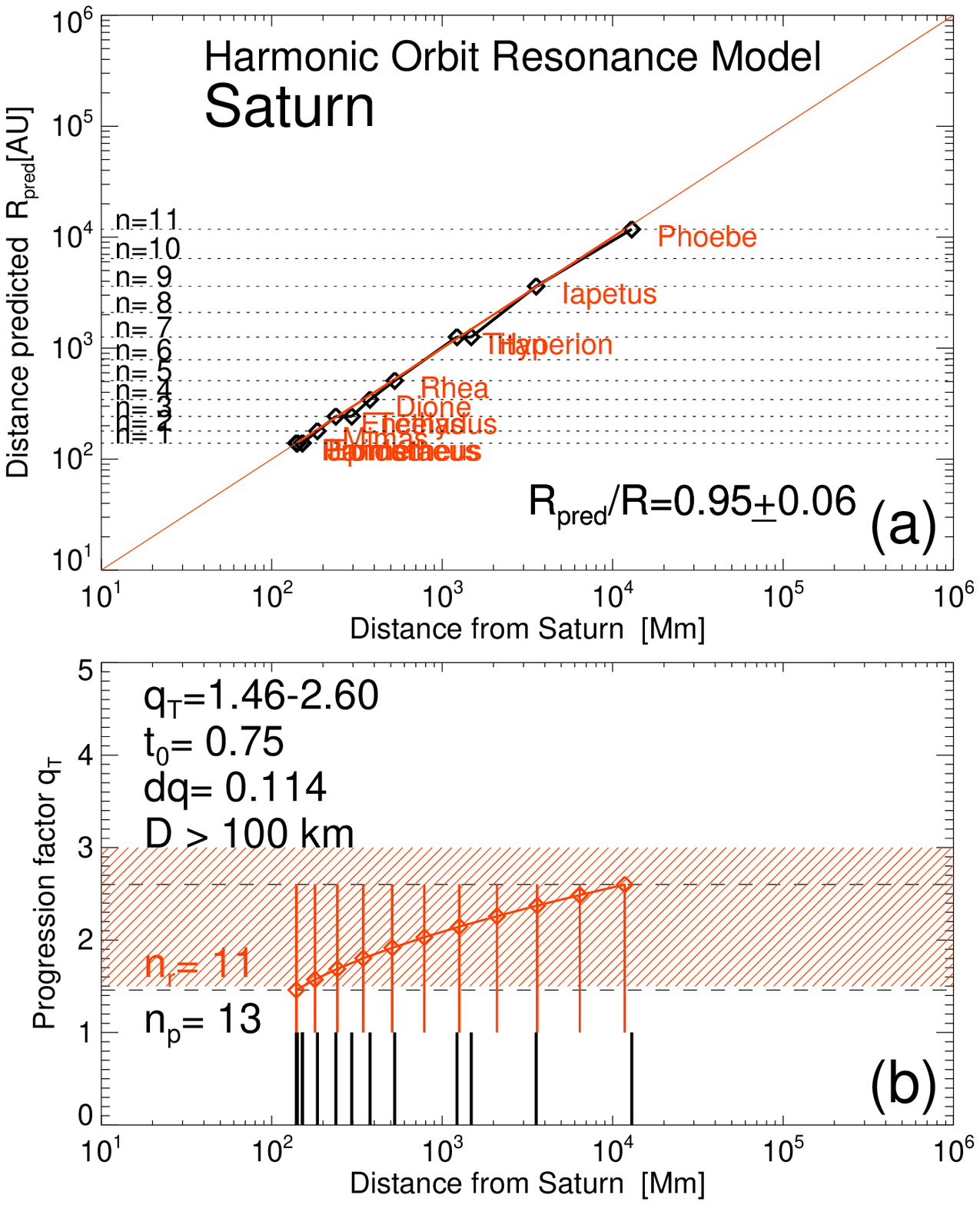}
\caption{The distances of the 11 largest ($D>100$ km) 
moons from Saturn are calculated with a harmonic
orbit resonance model (a), for which the progression factor
increases linearly with the orbit time (b). Representation
otherwise similar to Fig.~5. Note that $n=13$ resonant
zones fit $n=11$ observed moon distances.}
\end{figure}

\begin{figure}
\plotone{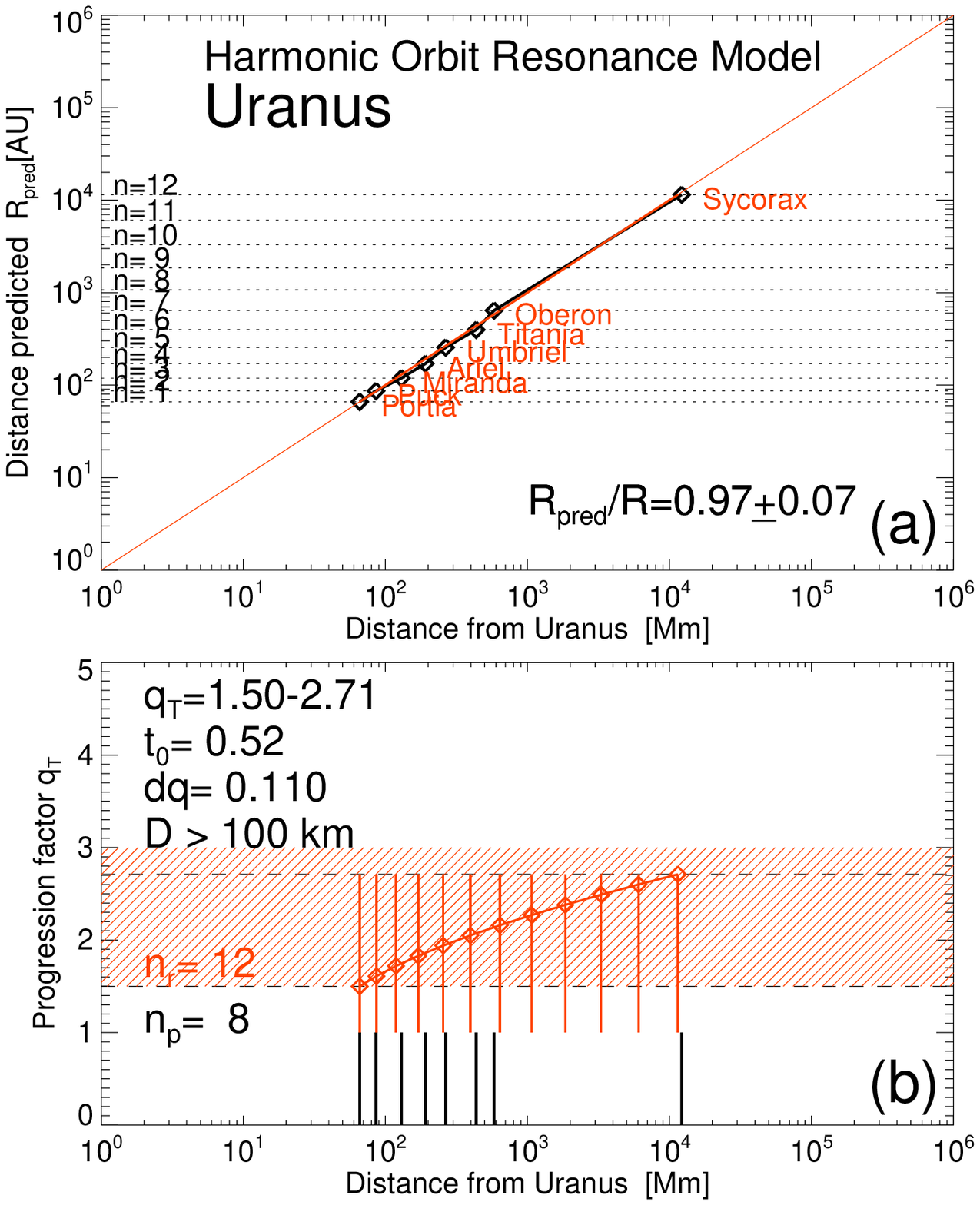}
\caption{The distances of the eight largest ($D>100$ km) 
moons from Uranus are calculated with a harmonic
orbit resonance model (a), for which the progression factor
increases linearly with the orbit time (b). Representation
otherwise similar to Fig.~5. Note that $n=12$ resonant
zones fit $n=8$ observed moon distances.}
\end{figure}

\begin{figure}
\plotone{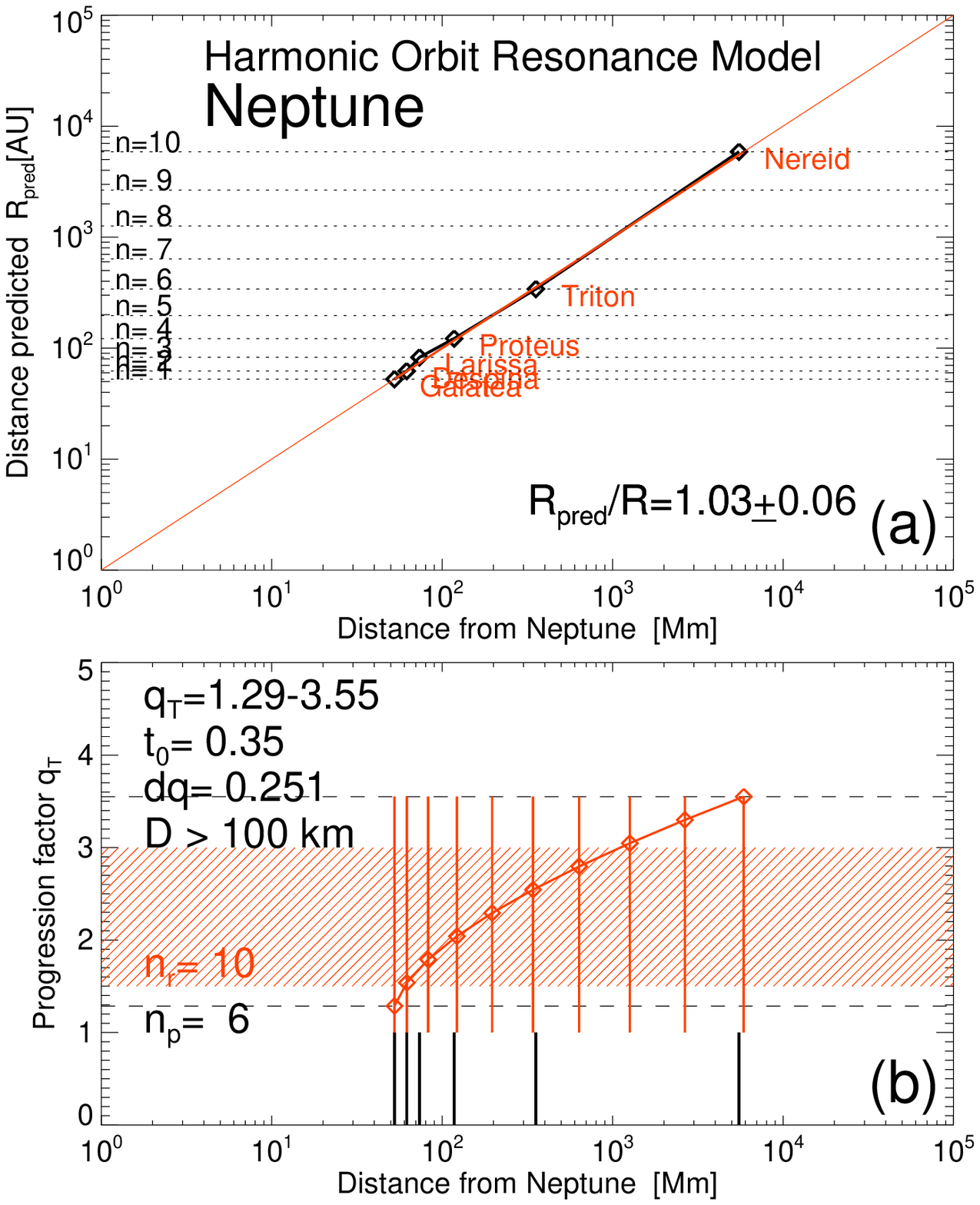}
\caption{The distances of the six largest ($D>100$ km) 
moons from Neptune are calculated with a harmonic
orbit resonance model (a), for which the progression factor
increases linearly with the orbit time (b). Representation
otherwise similar to Fig.~5. Note that $n=10$ resonant
zones fit $n=6$ observed moon distances.}
\end{figure}

\begin{figure}
\plotone{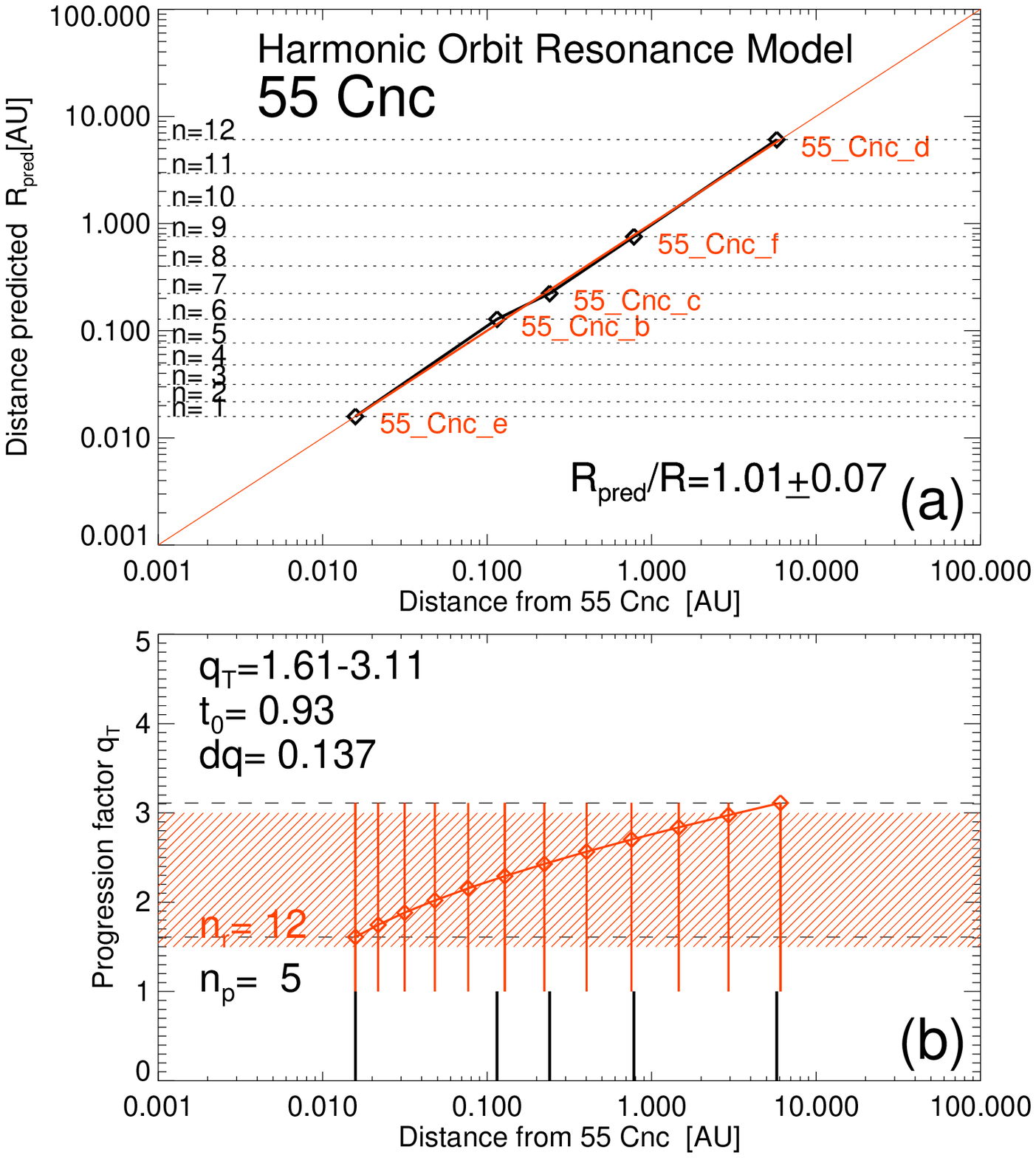}
\caption{The distances of 5 exo-planets from their central 
star 55 Cnc are calculated with a harmonic
orbit resonance model (a), for which the progression factor
increases linearly with the orbit time (b). Representation
otherwise similar to Fig.~5. Note that $n=12$ resonant
zones fit $n=5$ observed exo-planet distances.}
\end{figure}

\begin{figure}
\plotone{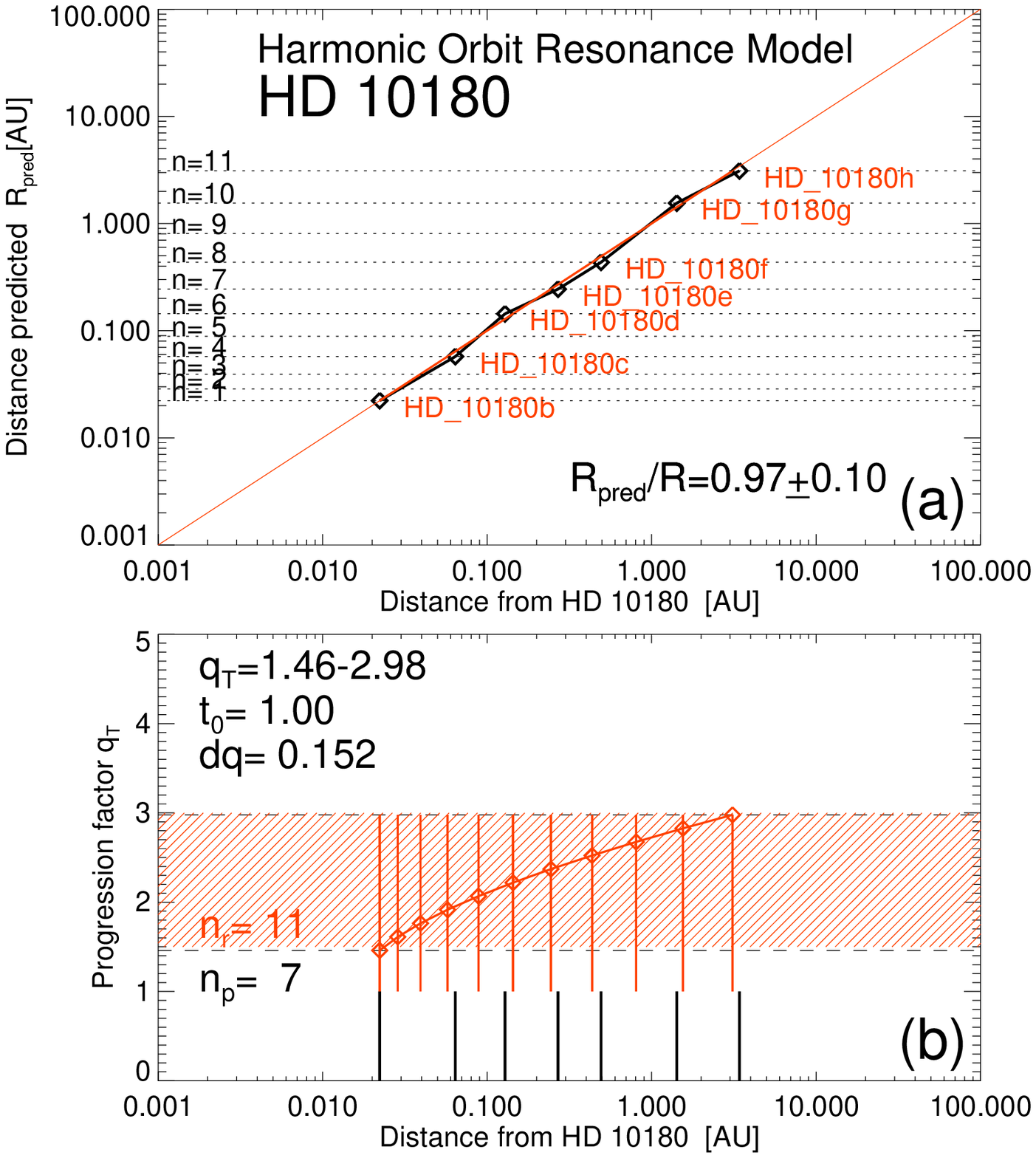}
\caption{The distances of 7 exo-planets from their central 
star HD 10180 are calculated with a harmonic
orbit resonance model (a), for which the progression factor
increases linearly with the orbit time (b). Representation
otherwise similar to Fig.~5. Note that $n=11$ resonant
zones fit $n=7$ observed exo-planet distances.}
\end{figure}

\begin{figure}
\plotone{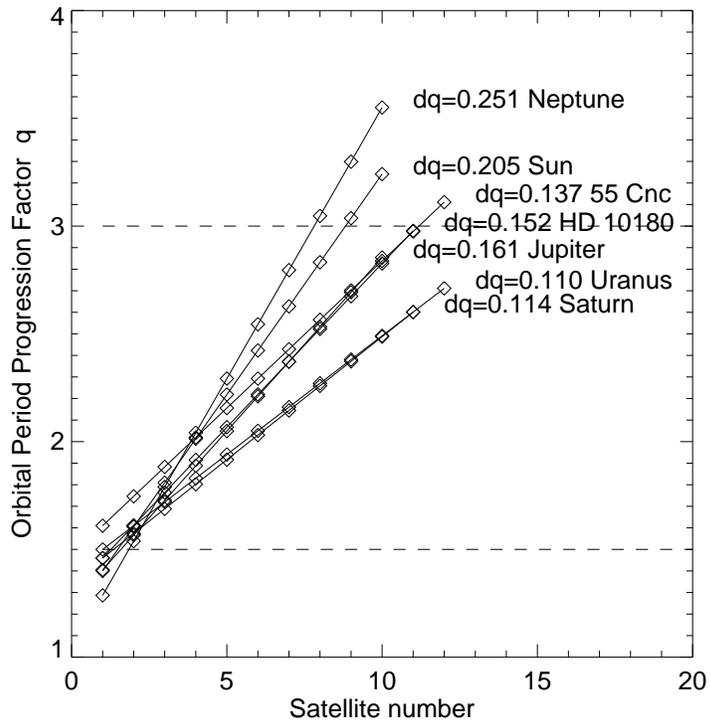}
\caption{The progression factor $q$ as a function of the
satellite number, obtained from the best fits of the harmonic
orbit resonance model to the data for the 8 data sets shown
in Figs.~4-11.}
\end{figure}

\begin{figure}
\plotone{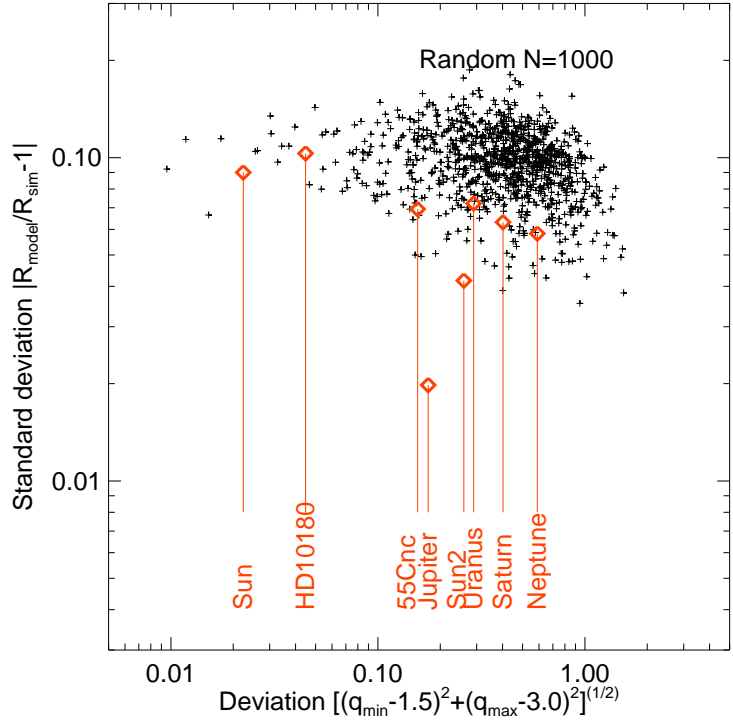}
\caption{1000 simulations (black crosses) of randomly distributed
planet distances $R$ are compared with the seven analyzed
data sets (red diamonds). The quantity on the y-axis represents
the standard deviation of the ratio of model distances $R_{model}$
to the simulated distances $R_{sim}$ for 1000 data sets with 10
planets each. The quantity on the x-axis represents the statistical
deviation of the best-fit $q$-values (of the time period
progression factor) from the theoretically predicted values
$q_{min}=1.5$ and $q_{max}=3.0$. Note that the results from the
observed data sets are significantly different from those 
of the simulated random data sets.}
\end{figure}

\end{document}